\documentclass[12pt]{article}
\usepackage[T1]{fontenc}
\usepackage{amsmath}
\usepackage{amsfonts}
\usepackage{graphicx}
\usepackage[mathscr]{euscript}
\usepackage{rotating}
\usepackage{pdflscape}
\usepackage[hmargin=2.2cm,vmargin=2.75cm]{geometry} 
\usepackage{booktabs,caption}
\usepackage{natbib}
\usepackage{setspace}
\usepackage{bbm}
\usepackage{amsthm}
\usepackage{enumerate}
\usepackage{threeparttable}
\usepackage{verbatim}
\usepackage{caption}
\usepackage{url}
\usepackage{amssymb}
\usepackage{subcaption,array}
\usepackage{adjustbox}
\usepackage[titletoc,title]{appendix}

\usepackage[hidelinks,breaklinks]{hyperref}
\hypersetup{colorlinks=true,allcolors=blue}

\usepackage{framed}
\usepackage[dvipsnames]{xcolor}

\theoremstyle{definition}
\newtheorem{assumption}{ASSUMPTION}

\DeclareMathOperator*{\argmax}{arg\,max}
\usepackage{calligra}
\usepackage[T1]{fontenc}

\begin{document}

\title{\textbf{Dynamic demand for differentiated products \\with fixed-effects unobserved heterogeneity}\thanks{I would like to thank helpful comments from the editor, Jaap Abbring, and two anonymous referees. I have benefited from discussions with Jiaying Gu, Matthew Osborne, Jes\'{u}s Carro, Bo Honor\'{e}, Francis Guiton, Lance Lochner, Nail Kashaev, Enrique Sentana, and 
Orazio Attanasio, and from comments from the audiences in the Applied Economics and Econometrics seminar at Western University, the special session on Econometrics of Dynamic Discrete Choice at the 2021 conference of the Royal Economic Society, and the 2022 Conference on econometric methods and empirical analysis of micro data in honor of Manuel Arellano at Banco de España.}}

\author{Victor Aguirregabiria\footnote{Department of Economics, University of Toronto. 150 St. George Street, Toronto, ON, M5S 3G7, Canada, \href{mailto: victor.aguirregabiria@utoronto.ca}{victor.aguirregabiria@utoronto.ca}.} \\ \emph{University of Toronto, CEPR}}

\date{August 15, 2022}

\maketitle

\thispagestyle{empty}

\begin{abstract}
This paper studies identification and estimation of a dynamic discrete choice model of demand for differentiated product using consumer-level panel data with few purchase events per consumer (i.e., short panel). Consumers are forward-looking and their preferences incorporate two sources of dynamics: \textit{last choice dependence} due to habits and switching costs, and \textit{duration dependence} due to inventory, depreciation, or learning. A key distinguishing feature of the model is that consumer unobserved heterogeneity has a \textit{Fixed Effects (FE)} structure -- that is, its probability distribution conditional on the initial values of endogenous state variables is unrestricted. I apply and extend recent results to establish the identification of all the structural parameters as long as the dataset includes four or more purchase events per household. The parameters can be estimated using a \textit{sufficient statistic - conditional maximum likelihood} (CML) method. An attractive feature of CML in this model is that the sufficient statistic controls for the forward-looking value of the consumer's decision problem such that the method does not require solving dynamic programming problems or calculating expected present values.

\vspace{0.4cm}

\vspace{0.4cm}
\end{abstract}

\newpage
\setcounter{page}{1}

\setstretch{1.5}


\section{Introduction\label{sec:intro}}

In many markets, consumer demand is dynamic in the sense that consumers' utility depends on past decisions. Sources of dynamics in demand include, among others, storable products,\footnote{See \citeauthor{boizot_robin_2001} (\citeyear{boizot_robin_2001}), \citeauthor{pesendorfer_2002} (\citeyear{pesendorfer_2002}), \citeauthor{erdem_imai_2003} (\citeyear{erdem_imai_2003}), and \citeauthor{hendel_nevo_ecma_2006} (\citeyear{hendel_nevo_ecma_2006}).} durable products,\footnote{See \citeauthor{esteban_shum_2007} (\citeyear{esteban_shum_2007}), \citeauthor{goettler_gordon_2011} (\citeyear{goettler_gordon_2011}), and \citeauthor{gowrisankaran_rysman_2012} (\citeyear{gowrisankaran_rysman_2012}).} habit formation and switching costs,\footnote{See \citeauthor{roy_chintagunta_1996} (\citeyear{roy_chintagunta_1996}), \citeauthor{keane_1997} (\citeyear{keane_1997}), and \citeauthor{osborne_2011} (\citeyear{osborne_2011}).} adoption costs,\footnote{See \citeauthor{ryan_tucker_2012} (\citeyear{ryan_tucker_2012}), and \citeauthor{degroote_verboven_2019} (\citeyear{degroote_verboven_2019}).} and learning.\footnote{See \citeauthor{ackerberg_2003} (\citeyear{ackerberg_2003}), and \citeauthor{ching_2010} (\citeyear{ching_2010}).} These models exhibit two main forms of \textit{state dependence} in consumers' purchasing decisions: \textit{last-choice dependence} and \textit{duration dependence}. We have  \textit{last-choice dependence} if a consumer's previous purchase of a product has a causal effect on her current probability of buying that product, for instance, as a result of habits, switching costs, or learning. We have \textit{duration dependence} if the time elapsed since the last purchase has a causal effect on the current buying decision. Depletion of storable products and depreciation of durable products generate duration dependence in consumer demand. These forms of \textit{state dependence} can induce substantial differences between short-run and long-run responses of demand to price changes. This has important economic implications in applications such as evaluating the effects of taxes and subsidies, measurement of firms' market power, or consumer welfare. The estimation of dynamic structural demand models using consumer panel data tries to measure these causal effects and use them for counterfactual analysis and welfare evaluation.\footnote{\citeauthor{nevo_2011} (\citeyear{nevo_2011}) and \citeauthor{berry_haile_2021} (\citeyear{berry_haile_2021}) are excellent surveys on identification and estimation of static models of demand for differentiated product. For surveys on dynamic models of consumer demand, see  \citeauthor{aguirregabiria_nevo_2013} (\citeyear{aguirregabiria_nevo_2013}), 
\citeauthor{keane_2015} (\citeyear{keane_2015}), and section 6.2 in \citeauthor{gandhi_nevo_2021} (\citeyear{gandhi_nevo_2021}).}

Unobserved heterogeneity plays a fundamental role in dynamic demand models using consumer panel data. Ignoring or incorrectly specifying the correlation between unobserved heterogeneity and pre-determined explanatory variables (e.g., previous purchasing decisions) can generate important biases in the estimation of the structural parameters that capture dynamic causal effects (\citeauthor{heckman_1981_statistical}, \citeyear{heckman_1981_statistical}). Furthermore, the distribution of consumer taste heterogeneity has an important impact on demand price elasticities (\citeauthor{berry_levinshon_1995}, \citeyear{berry_levinshon_1995}). The empirical literature on dynamic demand of differentiated product has considered a \textit{Random Effects (RE)} approach to model consumer unobserved heterogeneity. \textit{RE} models impose parametric restrictions on the distribution of unobserved heterogeneity and on the correlation between these unobservables and the initial values of the predetermined explanatory variables.\footnote{In the literature of dynamic discrete choice structural models, the most common approach to deal with time-invariant unobserved heterogeneity is assuming that it has a finite mixture distribution with a small number of points of support – often as small as two or three unobserved types. The probability of the initial values of the state variables is unrestricted conditional on an unobserved individual type. Some studies allow the distribution of unobserved types and the probability of initial conditions varies with observable time-invariant individual characteristics. A few studies assume that the probability of the initial conditions comes from the ergodic distribution of the state variables. For more details, see the survey papers by \citeauthor{aguirregabiria_mira_2010} (\citeyear{aguirregabiria_mira_2010}), \citeauthor{arcidiacono_ellickson_2011} (\citeyear{arcidiacono_ellickson_2011}), or \citeauthor{keane_2015} (\citeyear{keane_2015}).} Though this distribution is not identifiable in short panels, its misspecification can generate important biases in the estimation of dynamic causal effects. This is the so called \textit{initial conditions problem} (\citeauthor{heckman_1981_incidental}, \citeyear{heckman_1981_incidental}). Therefore, \textit{RE} models are not robust to misspecification of parametric restrictions on unobserved heterogeneity.\footnote{In the context of dynamic discrete choice demand models, a recent paper by \citeauthor{simonov_dube_2020} (\citeyear{simonov_dube_2020}) shows that the bias induced by the misspecification of the probability of the initial conditions can be sizable. The authors show that 
Random Effects specifications commonly used in empirical applications, if misspecified, can lead to biases larger than two times the true value of the structural parameter. This magnitude of the bias appears even in relatively long panel data sets spanning more than three or four years of shopping history.} \textit{Fixed effects (FE)} approaches impose no restriction on the distribution of consumers' unobserved heterogeneity such that the identification of parameters of interest is more robust than in \textit{RE} models. However, several identification concerns have inclined researchers to avoid a \textit{FE} approach in applications of dynamic discrete choice structural models, and more specifically, in dynamic demand models.\footnote{An interesting exception is \citeauthor{jones_landwehr_1988} (\citeyear{jones_landwehr_1988}). The authors estimate a Fixed Effects dynamic binary logit model for consumers' decision of buying the same brand as in last purchase or switching to a different brand. They estimate this model using a conditional maximum likelihood method. In their model, consumers are not forward-looking, there is no duration dependence, there are no product characteristics (e.g., prices), and it is a binary choice model.}

A first issue is related to the identification of structural parameters. All the existing (positive) identification results in \textit{FE} dynamic discrete choice models impose the restriction that unobserved heterogeneity enters additively in the utility of each choice alternative. However, in dynamic programming models, unobserved heterogeneity enters not only in current utility but also in the continuation value of the forward-looking decision problem, and these continuation values depend non-additively (and in fact, without a closed-form expression) on both unobserved heterogeneity and observable state variables. The common wisdom was that \textit{FE} models cannot deal with the non-additive unobserved heterogeneity that is inherent to discrete choice dynamic programming models.  
A second important issue is related to the fact that \textit{FE} methods cannot deliver identification of the distribution of unobserved heterogeneity with short-panels. Most empirical applications of dynamic structural models are interested in using the estimated model to obtain \textit{Average Marginal Effects (AME)} on endogenous variables of changes in explanatory variables or in structural parameters. Demand price elasticities are examples of these \textit{AME}s. Until very recently, the common wisdom was that \textit{AME}s are not identified in \textit{FE} models, as they are expectations over the distribution of the unobserved heterogeneity (e.g., \citeauthor{abrevaya_hsu_2021}, \citeyear{abrevaya_hsu_2021}; \citeauthor{honore_depaula_2021}, \citeyear{honore_depaula_2021}). 

Two recent studies provide positive identification results for the two issues discussed above.  \citeauthor{aguirregabiria_gu_2021} (\citeyear{aguirregabiria_gu_2021}) establish the identification of structural parameters in a class of \textit{FE} dynamic panel data logit models where agents are forward-looking. The class of models includes two types of endogenous state variables: the lagged choice variable, and the time duration in the last choice. \citeauthor{aguirregabiria_carro_2021} (\citeyear{aguirregabiria_carro_2021}) prove point identification of different AMEs in \textit{FE} dynamic logit models. For instance, the average causal effect of changes in the lagged dependent variable or in the duration in last choice are identified.\footnote{In related work, \citeauthor{chernozhukov_fernandez_2013} (\citeyear{chernozhukov_fernandez_2013}), and more recently \citeauthor{davezies_2021} (\citeyear{davezies_2021}), provide partial identification results for a general class of AMEs in binary choice panel data models.}

In this paper, I apply and extend identification results in \citeauthor{aguirregabiria_gu_2021} (\citeyear{aguirregabiria_gu_2021}) to a \textit{FE} dynamic panel data model of consumer demand with differentiated products. The model can incorporate storable or durable products, habit formation, and brand switching costs. I present a new identification result in FE dynamic discrete choice with forward-looking agents. I show identification of utility parameters associated to state variables that follow exogenous stochastic processes. More specifically, in the context of demand models, I establish the identification of parameters that capture the effect of prices. In a FE forward-looking model, this identification result relies on a particular structure in the stochastic process of prices. There are two components in prices: a persistent component and a transitory component (i.e., temporary promotion).

The structural parameters of the model are estimated using a \textit{sufficient statistics - conditional maximum likelihood (CML)} method for dynamic discrete choice panel data models. Seminal work on this method is due to \citeauthor{cox_1958} (\citeyear{cox_1958}), \citeauthor{rasch_1960} (\citeyear{rasch_1960}, \citeyear{rasch_1961}), and \citeauthor{andersen_1970} (\citeyear{andersen_1970}), and in econometrics by \citeauthor{chamberlain_1985} (\citeyear{chamberlain_1985}). More specifically, I apply the kernel-weighted method in \citeauthor{honore_kyriazidou_2000} (\citeyear{honore_kyriazidou_2000}). In the context of dynamic structural models, a very helpful implication of this CML method is that it is not subject to a \textit{curse of dimensionality} associated with the computational cost of solving a dynamic programming problem, or calculating present values of future utilities. This is particularly relevant in applications of dynamic demand of differentiated product because the dimension of the state space increases exponentially with the number of products, which is typically large. The \textit{sufficient statistics - CML} method "differences out" the continuation (forward-looking) value of the consumer's decision problem, as this value depends on the incidental parameters / unobserved heterogeneity. This implies that continuation values need not be computed to implement the CML estimator. Therefore, the CML estimation of this dynamic structural model is computationally as simple as in static or myopic models, and its cost does not depend on the dimension of the state space.\footnote{CML estimation implies a different type of curse of dimensionality. The evaluation of the conditional likelihood function involves adding up a function over all the possible values of the vector of sufficient statistics. The number of possible values increases exponentially with the number of time periods in the data, $T$. However, this dimensionality problem has an easy solution -- at the cost of some efficiency lost in the CML estimator -- which consists of splitting a $T$-periods history into shorter sub-histories.}

Related to the curse of dimensionality in the estimation of dynamic structural models, it is worth to note that the CML method in this paper is quite different to the \textit{Finite Dependence -- Conditional Choice Probabilities (FD-CCP)} approach in \citeauthor{arcidiacono_miller_2011} (\citeyear{arcidiacono_miller_2011}). The FD-CCP approach is a Random Effects method that requires parametric assumptions on the distribution of the unobserved heterogeneity, typically, a finite mixture structure. The FD-CCP method also requires consistent nonparametric estimates of conditional choice probabilities for each unobserved individual type. Consistent nonparametric estimates of type-specific CCPs are available -- under some conditions -- in Random Effects finite mixture models (see \citeauthor{kasahara_shimotsu_2009}, \citeyear{kasahara_shimotsu_2009}), but not in Fixed Effects models. 

In an influential paper, \citeauthor{hendel_nevo_ecma_2006} (\citeyear{hendel_nevo_ecma_2006}) propose an approach to deal with the curse of dimensionality in the estimation of structural dynamic discrete choice models of consumer demand. Under the restrictions of (i) no brand switching costs, (ii) no product differentiation in depreciation / depletion rates, and (iii) no product differentiation in consumption (only at the moment of purchase), a consumer brand choice decision is a static decision, and the only dynamic decision is the replenishment or (repurchasing time) decision. This structure – together with the behavioral assumption that consumers believe that McFadden’s surplus (i.e., inclusive value) follows a Markov process -- facilitates substantially the estimation of the dynamic structural model. In contrast, in this paper, I consider a model that includes brand switching costs, product differentiation in depletion rates, and product differentiation in consumption, so that consumer brand choice is dynamic. Furthermore, the sufficient statistic result in my paper relies on the structure of the utility function, but it does not exploit any additional behavioral restriction on consumers’ beliefs.

In a recent paper, \citeauthor{kong_dube_2022} (\citeyear{kong_dube_2022}) apply the \textit{Dynamic Potential Outcomes (DPO)} model in \citeauthor{torgovitsky_2019} (\citeyear{torgovitsky_2019}) to study the identification of the causal effect of consumers’ past purchasing decisions on current choices, and more specifically the separation between brand switching costs and unobserved heterogeneity. The DPO model can be interpreted as the reduced form of a dynamic structural model that allows for a very general form of consumer unobserved heterogeneity including, for instance, unobserved heterogeneity in consumer switching costs. Using the DPO model and consumer panel data -- and in the absence of a controlled randomized experiment -- the causal effect of state dependence is not point identified. The authors characterized the identified set for the state dependence generated by brand switching costs. They also show that consumer forward-looking behavior in the presence of brand switching costs implies exclusion restrictions that satisfy the conditions in \citeauthor{abbring_daljord_2020} (\citeyear{abbring_daljord_2020}) to identify the discount factor and the utility function jointly.

Motivated by the difficulty of estimating Random Effects models with rich forms of unobserved heterogeneity, \citeauthor{chesher_santossilva_2002} (\citeyear{chesher_santossilva_2002}) propose an extension of the standard multinomial logit model and a simple two-step estimation method that shares some similarities with the sufficient-statistics conditional likelihood approach in this paper. Though the authors present their model/method in a static / cross-sectional context, it seems that it can be applied to a dynamic logit model and combined with the method in this paper to allow for unobserved heterogeneity in the structural parameters capturing state dependence.

The rest of the paper is organized as follows. Section \ref{sec:model} describes the model. Section \ref{sec_identification} presents identification results. Section \ref{sec_estimation} describes the estimation method. In section \ref{sec_conclusions}, I summarize and discuss some extensions for further research.


\section{Model \label{sec:model}}

I present a framework that includes both storable and durable differentiated products.  Most of the features of the model are standard in the literature. The main distinguishing feature is the \textit{Fixed Effects (FE)} nature of consumer unobserved heterogeneity.

\subsection{Basics}

There are $J$ different brands of a differentiated product, and we index brands by $j \in \{1,2,...,J\}$.  We index consumers by $i$. Time is discrete and indexed by $t$. There is a dichotomy in the definition of \textit{time} in empirical applications of dynamic consumer demand. In most applications, $t$ has the standard interpretation as calendar time: for instance, the unit of time can be a week. In these applications, the model and data account for time periods where a consumer does not purchase any variety of the differentiated product. In contrast, a good number of empirical applications in this literature consider $t$ as the index for purchase events, such that $t=1$ means a consumer's first purchase, $t=2$ is her second purchase, and so on (see \citeauthor{keane_1997}, \citeyear{keane_1997}; \citeauthor{osborne_2011}, \citeyear{osborne_2011}; among others). The different definition of time $t$ has implications on the interpretation of dynamics in the model, such as duration dependence.\footnote{Also very importantly, the model where $t$ indexes purchase events does not include the decision of "no purchase" as a choice alternative. That is, there is not an \textit{outside alternative}.} Here, I follow the standard definition of $t$ as calendar time. However, all the identification results in this paper apply also to models where $t$ indexes purchase events.

Every period $t$, consumer $i$ decides whether to purchase or not one unit of the product, and which brand to buy. Variable $y_{it} \in \mathcal{Y}=\{0,1,...,J\}$ represents the decision of consumer $i$ at period $t$, where $y_{it} = 0$ means "no purchase", and $y_{it} = j>0$ means purchase of brand $j$. There are two endogenous state variables that capture dynamics in consumer demand: brand choice in the last purchase, $\ell_{it} \in \{1, 2, ..., J\}$; and time duration since the last purchase, $d_{it} \in \{1, 2, ..., \infty\}$.\footnote{At this point, I consider that the space of the duration variable is unbounded from above. In Section 3 below, I introduce \textcolor{blue}{Assumption} \ref{assumption_3} that implies that there is a finite value, $d^{*}$, for the duration variable such that a consumer’s utility function does not change for values of duration greater than $d^{*}$. Under this \textcolor{blue}{Assumption} \ref{assumption_3}, we can restrict the space of the duration variable to be $\{1,2, ..., d^{*}\}$. Importantly, the value $d^{*}$ is identified from the data as long as $d^{*} \leq (T-1)/2$, where $T$ is the number of time periods in the data.} Last brand choice is related to habit formation and switching costs. Duration since last purchase captures the effect of inventory depletion (for storable products) or depreciation (for durable products). By definition, the transition rule of the vector of endogenous state variables $\mathbf{x}_{it} \equiv (\ell_{it}, d_{it})$ is:
\begin{equation}
    \mathbf{x}_{i,t+1} \equiv 
    \left( \ell_{i,t+1} \text{ } , \text{ } d_{i,t+1} \right)
    \text{ } = \text{ }
    f_{x}(y_{it},\mathbf{x}_{it})
    \text{ } \equiv \text{ }
    \left\{
    \begin{array}
    [c]{rcl}
    \left( \ell_{it} \text{ } , \text{ } d_{it}+1 \right)
    & if & y_{it}=0
    \\
    \left( j \text{ } , \text{ } 1 \right)
    & if & y_{it}=j>0
    \end{array}
    \right. 
\label{eq_transition_endogenous}
\end{equation}

In many applications, consumers are located in different geographic markets and face different prices. This is not a necessary condition for the identification results in this paper, but I allow for it. Let $p_{it}(j)$ be the price of product $j$ at period $t$ in the market where consumer $i$ is located, and let $\mathbf{p}_{it}$ be the vector with the prices of all products, $\mathbf{p}_{it} \equiv (p_{it}(j):j=1,2,...,J)$.

\subsection{Utility \label{sec:model_utility}}

Let $U_{it}$ be the per-period utility that consumer $i$ obtains at period $t$. It has four components:
\begin{equation}
    U_{it} \text{ } = \text{ }
    b_{i} \left( y_{it}, \ell_{it}, d_{it} \right)
    \text{ } + \text{ } 
    m_{i} \left( y_{it}, \mathbf{p}_{it} \right)
    \text{ } - \text{ }
    sc_{i} \left( y_{it}, \ell_{it} \right)
    \text{ } + \text{ }
    \varepsilon_{it}\left( y_{it} \right)
\label{eq_perperiod_utility}
\end{equation}
The term $b_{i} \left( y_{it}, \ell_{it}, d_{it} \right)$ is the utility from consumption of the branded product; $m_{i} \left( y_{it}, \mathbf{p}_{it} \right)$ represents utility from consumption of the outside good or numeraire;  $sc_{i} \left( y_{it}, \ell_{it} \right)$ captures brand switching costs or habits; and $\varepsilon_{it}\left( y_{it} \right)$ is a consumer idiosyncratic taste shock at the moment of purchase. I describe below each of these components.

\bigskip

\noindent \textbf{(i) Utility from consumption of the branded product.} The flow utility from consumption of the branded product 
has the following form:
\begin{equation}
    b_{i} \left( y_{it}, \ell_{it}, d_{it} \right) 
    \text{ } \equiv \text{ }
    \left\{
    \begin{array}
    [c]{lcl}
        \alpha_{i}(j) 
        & if & y_{it}=j>0
    \\
    \alpha_{i}(\ell_{it}) - 
    \beta^{dep}(\mathbf{w}_{i},\ell_{it}) \text{ } d_{it}
    & if & y_{it}=0 \text{ }.
    \end{array}
    \right. 
\label{eq_utility_branded_product_2}
\end{equation}

The model allows for product differentiation at the moment of consumption and not only at the moment of purchase. Parameter $\alpha_{i}(j)$ is the flow utility that consumer $i$ receives from consuming product $j$ when the product has not depreciated (or depleted) over time. The vector $\boldsymbol{\alpha}_{i} \equiv (\alpha_{i}(1), \alpha_{i}(2), ..., \alpha_{i}(J))$ represents the fixed effects for consumer $i$. The fixed effect $\alpha_{i}(j)$ depends on product and consumer characteristics in an unrestricted way. For instance, it could have the structure $\alpha_{i}(j) = \boldsymbol{x}_{j}^{\prime}\boldsymbol{\beta}_{i} + 
\boldsymbol{\xi}_{j}^{\prime}\boldsymbol{\omega}_{i}$, where 
$\boldsymbol{x}_{j}$ and $\boldsymbol{\xi}_{j}$ are vectors of observable and unobservable product characteristics, respectively, and $\boldsymbol{\beta}_{i}$ and $\boldsymbol{\omega}_{i}$ are the vectors of marginal utilities of these characteristics for consumer $i$. Therefore, the specification of consumer heterogeneity in preferences is very flexible.

The flow utility from consumption of the branded product takes into account depletion or depreciation over time.\footnote{The datasets used in this literature contain information on consumer purchase histories, but not on consumption and inventories of storable products. To deal with this missing information for storable products, researchers have imposed different restrictions. \citeauthor{aguirregabiria_nevo_2013} (\citeyear{aguirregabiria_nevo_2013}) discuss this issue and different restrictions in the literature. A common approach is using duration since last purchase, $d_{it}$, as a proxy of inventory. For storable products, this is the approach in \citeauthor{erdem_imai_2003} (\citeyear{erdem_imai_2003}).} If the consumer purchases a new product (if $y_{it}=j>0$), there is no depreciation effect and the flow utility is $\alpha_{i}(j)$. Otherwise, if the consumer does not purchase a new product (if $y_{it}=0$), she consumes product $\ell_{it}$ -- her last purchase -- and flow utility is $\alpha_{i}(\ell_{it}) - \beta^{dep}(\mathbf{w}_{i},\ell_{it}) d_{it}$, where $\mathbf{w}_{i}$ is a vector of consumer characteristics observable to the researcher, e.g., income, education, geographic location. Parameter (or function) $\beta^{dep}(\mathbf{w}_{i},j)$ represents the depreciation (depletion) rate in product $j$. This depreciation/depletion effect can vary across products.

\bigskip

\noindent \textbf{(ii) Utility from consumption of the composite good.} The consumer’s budget constraint implies that consumption of the composite good is equal to $\mu_{i} - \sum_{j=1}^{J} p_{it}(j) \text{ } 1\{y_{it}=j\}$, where 
where $\mu_{i}$ is the consumer's (weekly) disposable income, and $\sum_{j=1}^{J} p_{it}(j) \text{ } 1\{y_{it}=j\}$ represents expenditure in the branded product. The utility from consumption of the composite good is:
\begin{equation}
    m_{i} \left( y_{it}, \mathbf{p}_{it} \right) 
    \text{ } = \text{ }
    \gamma \left( \mathbf{w}_{i} \right)
    \text{ }
    h\bigg(
        \mu_{i} - 
        \sum_{j=1}^{J} p_{it}(j) \text{ } 1\{y_{it}=j\}
    \bigg),
\label{eq_budget_constraint}
\end{equation}
where $h(c)$ is a strictly increasing function that is known to the researcher, e.g. linear ($h(c)=c$), logarithmic ($h(c)=ln(c)$); and $\gamma \left( \mathbf{w}_{i} \right)$ is a structural parameter (function) related to the marginal utility of consuming the composite good and may depend on consumer observable characteristics $\mathbf{w}_{i}$. When function $h(c)$ is nonlinear -- e.g., logarithmic -- the identification results in this paper require that the researcher observes consumer disposable income $\mu_{i}$. For the rest of the paper, I assume that this is the case.

\bigskip

\noindent \textbf{(iii) Product switching costs / habits.} Following the standard approach, brand switching costs take place at the moment of purchase, and not through the consumption of the product. The specification of brand switching costs is:
\begin{equation}
    sc_{i} \left( y_{it}, \ell_{it} \right) 
    \text{ } = \text{ }
    \beta^{sc}(\mathbf{w}_{i},\ell_{it}, y_{it}),
\label{eq_switching_costs}
\end{equation}
Parameter $ \beta^{sc}(\mathbf{w}_{i},k,j)$ represents the cost of switching from brand $k$ to brand $j$. There is no switching cost without switching or without a new purchase, such that such that $\beta^{sc}(\mathbf{w}_{i},k,k)=0$ and $\beta^{sc}(\mathbf{w}_{i},k,0)=0$. 

This component of the utility can be also interpreted in terms of \textit{habit formation}. In that case, we have that $sc_{i} \left( y_{it}, \ell_{it} \right) = -\beta^{hab}(\mathbf{w}_{i},\ell_{it}, y_{it})$, where parameter $\beta^{hab}(\mathbf{w}_{i},j,j) \geq 0$ for $j>0$ is the increase in utility from purchasing the same brand as in last purchase, and for any $k \neq j$, $\beta^{hab}(\mathbf{w}_{i},k,j)$ is restricted to be zero. This \textit{habit formation} model is equivalent to the \textit{brand switching cost} model under the restrictions $\beta^{sc}(\mathbf{w}_{i},k,j) =  \beta^{sc}(\mathbf{w}_{i},k',j)$ for any $k,k' \neq j$. This more parsimonious specification is quite common in this literature.

\citeauthor{keane_1997} (\citeyear{keane_1997}, \citeyear{keane_2015}) and
\citeauthor{bronnenberg_dube_2017} (\citeyear{bronnenberg_dube_2017}) discuss different specifications of brand loyalty / habit formation. The dependence of utility only on the last purchase is a limited form of habit formation. However, it has been used in a good number of influential studies on brand loyalty such as \citeauthor{jones_landwehr_1988} (\citeyear{jones_landwehr_1988}), \citeauthor{seetharaman_ainslie_1999} (\citeyear{seetharaman_ainslie_1999}), \citeauthor{shum_2004} (\citeyear{shum_2004}), \citeauthor{osborne_2011} (\citeyear{osborne_2011}), \citeauthor{pavlidis_2017} (\citeyear{pavlidis_2017}), \citeauthor{simonov_dube_2020} (\citeyear{simonov_dube_2020}), \citeauthor{mysliwski_sanches_2020} (\citeyear{mysliwski_sanches_2020}), among others. There are simple extensions of this model that provide more flexible forms of habit formation. For instance, the utility of buying product $j$ may depend on the number of times the product has been chosen during the last $n$ purchase events, where $n$ is larger than one. More generally, following the seminal work by \citeauthor{guadagni_little_1983} (\citeyear{guadagni_little_1983}), the model can include a habits (brand loyalty) stock variable for each product, where the stock depreciates with a certain rate and increases when the product is purchased. Establishing identification of this, more flexible, model of habit formation in this Fixed Effects framework is an interesting topic for further research.

\bigskip

\noindent \textbf{(iv) Extreme value distributed shocks in preferences.} Finally, variable $\varepsilon_{it}(j)$ captures other consumer idiosyncratic factors affecting the utility from purchasing product $j$ at period $t$. Variables $\boldsymbol{\varepsilon}_{it} \equiv 
(\varepsilon_{it}(0)$, ..., $\varepsilon_{it}(J))$ are mean-zero i.i.d. over $(i,t,j)$ with a type I Extreme Value distribution. Both $\boldsymbol{\alpha}_{i}$ and $\boldsymbol{\varepsilon}_{it}$ are unobservable to the econometrician.

\bigskip

Putting together the different components of the utility function, we have:
\begin{equation}
    U_{it} = 
    \left\{
    \begin{array}[c]{lcl}
    \alpha_{i}(\ell_{it}) + \gamma \left( \mathbf{w}_{i} \right) 
    h \left(\mu_{i} \right) - 
    \beta^{dep}(\mathbf{w}_{i},\ell_{it}) \text{ } d_{it} +
    \varepsilon_{it}(0)
    & if & y_{it}=0
    \\
    \\
    \alpha_{i}(j) + \gamma \left( \mathbf{w}_{i} \right)
    h \left( \mu_{i} - p_{it}(j)  \right) - 
     \beta^{sc}(\mathbf{w}_{i},\ell_{it},j) +
    \varepsilon_{it}(j)
    & if & y_{it}=j>0,  
    \end{array}
    \right. 
\label{eq_total_utility}
\end{equation}
For the rest of the paper, I use function $u_{\boldsymbol{\alpha}_{i}}(y_{it},\mathbf{x}_{it},\mathbf{p}_{it})$ to represent the part of the utility that does not include the unobservable logit shocks, such that $U_{it} = u_{\boldsymbol{\alpha}_{i}}(y_{it},\mathbf{x}_{it},\mathbf{p}_{it}) + 
\varepsilon_{it}(y_{it})$.

\bigskip

\noindent \textbf{(v) A remark on the identification / normalization of switching costs parameters.} In discrete choice models, we can only identify parameters affecting the difference between the utilities of two choice alternatives, for instance, the utility differences $U_{it}(j) - U_{it}(0)$ for $j=1,2, ..., J$. For the model in this paper, this implies that we need some “normalization” restrictions of the switching cost parameters. These restrictions are needed even in the hypothetical (ideal) case where there is no time-invariant consumer unobserved heterogeneity, consumers are not forward-looking, and the researcher observes utility differences $U_{it}(j) - U_{it}(0)$ for $j=1,2, ..., J$. In this model, these utility differences have the following form:
\begin{equation}
    U_{it}(j) - U_{it}(0) = 
    \alpha_{i}(j) - \alpha_{i}(\ell_{it}) +
    \gamma \left[ h_{it}(j) - h_{it}(0) \right] -
    \beta^{sc}(\ell_{it},j) + \beta^{dep}(\ell_{it}) d_{it} +
    \varepsilon_{it}(j) - \varepsilon_{it}(0)
\label{eq_utility_differences}
\end{equation}
where $h_{it}(j) \equiv h(\mu_{i}-p_{it}(j))$ and $h_{it}(0) \equiv h(\mu_{i})$. 
To illustrate this issue, consider the simpler case with: $J=2$ products; $\alpha_{i}(j)=\alpha(j)$ for every consumer $i$; and $\gamma=0$. Conditional on duration $d_{it}=d$, we have only four possible utility differences: $U_{it}(1)-U_{it}(0)$ and $U_{it}(2)-U_{it}(0)$ for $l_{it}=1,2$. The mean values of these utility differences are:
\begin{equation}
    \left\{
    \begin{array}
    [c]{lcl}
    \mathbb{E} \left[ 
        U_{it}(1) - U_{it}(0) | \ell_{it}=1, d_{it}=d
    \right] 
    & = &
    \alpha(1) - \alpha(1) -
    \beta^{sc}(1,1) + \beta^{dep}(1) \text{ } d
    \\
    \mathbb{E} \left[ 
        U_{it}(2) - U_{it}(0) | \ell_{it}=1, d_{it}=d
    \right] 
    & = &
    \alpha(2) - \alpha(1) -
    \beta^{sc}(1,2) + \beta^{dep}(1) \text{ } d
    \\
    \mathbb{E} \left[ 
        U_{it}(1) - U_{it}(0) | \ell_{it}=2, d_{it}=d
    \right] 
    & = &
    \alpha(1) - \alpha(2) -
    \beta^{sc}(2,1) + \beta^{dep}(2) \text{ } d
    \\
    \mathbb{E} \left[ 
        U_{it}(2) - U_{it}(0) | \ell_{it}=2, d_{it}=d
    \right] 
    & = &
    \alpha(2) - \alpha(2) -
    \beta^{sc}(2,2) + \beta^{dep}(2) \text{ } d
    \end{array}
    \right. 
\label{eq_system_four_differences}
\end{equation}
The system of equations in \eqref{eq_system_four_differences} includes four restrictions and eight parameters: $\alpha(1)$, $\alpha(2)$, $\beta^{sc}(1,1)$, $\beta^{sc}(1,2)$, $\beta^{sc}(2,1)$, $\beta^{sc}(2,2)$, $\beta^{dep}(1)$, and $\beta^{dep}(2)$. We need additional restrictions. As mentioned above, if we interpret $\beta^{sc}(k,j)$ as a switching cost, it seems natural to assume that there is no switching cost without switching, so that $\beta^{sc}(1,1)=\beta^{sc}(2,2)=0$. These additional restrictions imply the identification of the depletion parameters: $\beta^{dep}(j) \text{ } d = \mathbb{E} \left[ U_{it}(j) - U_{it}(0) | \ell_{it}=j, d_{it}=d \right]$. However, these restrictions are not sufficient to identify the switching cost parameters $\beta^{sc}(1,2)$ and $\beta^{sc}(2,1)$ separately from the difference in the intercepts $\alpha(2) - \alpha(1)$. Nevertheless, it is straightforward to show that we can identify the following linear combination of switching cost parameters:
\begin{equation}
    \begin{array}
    [c]{l}
    \beta^{sc}(1,2) + \beta^{sc}(2,1) - \beta^{sc}(1,1) - \beta^{sc}(2,2) =
    \\
    - \mathbb{E} \left[ 
        U_{it}(2) - U_{it}(0) | \ell_{it}=1, d_{it}=d
    \right] 
    - \mathbb{E} \left[ 
        U_{it}(1) - U_{it}(0) | \ell_{it}=2, d_{it}=d
    \right] 
    \\
    + \mathbb{E} \left[ 
        U_{it}(1) - U_{it}(0) | \ell_{it}=1, d_{it}=d
    \right] 
    + \mathbb{E} \left[ 
        U_{it}(2) - U_{it}(0) | \ell_{it}=2, d_{it}=d
    \right] 
    \end{array}
\label{eq_identification_example}
\end{equation}
This result extends to a model with any number of products $J \geq 2$. Therefore, for the rest of the paper, I focus on the identification of parameters $\gamma(\mathbf{w}_{i})$, $\beta^{dep}(\mathbf{w}_{i},j)$, and the following linear combination of switching cost parameters:
\begin{equation}
    \widetilde{\beta}^{sc}(\mathbf{w}_{i},k,j) 
    \text{ } = \text{ }
    \beta^{sc}(\mathbf{w}_{i},k,j) + \beta^{sc}(\mathbf{w}_{i},j,k) - \beta^{sc}(\mathbf{w}_{i},k,k) - \beta^{sc}(\mathbf{w}_{i},j,j)
\label{eq_betasc_tilde}
\end{equation}

\subsection{Stochastic process for prices}

The structure of the stochastic process of prices that I describe here is not necessary for the identification of the parameters $\beta^{sc}$ and $\beta^{dep}$, but it plays an important role in the identification of the marginal utility of income parameter, $\gamma$, in this forward-looking model. 

\bigskip

\begin{assumption}
    \textit{The price of any product $j$ has two components, one that is persistent over time ($z_{it}(j)$), and other that is transitory ($e_{it}(j)$).\footnote{Variables $z_{it}(j)$ and $e_{it}(j)$ can be scalars or vectors. For instance, in Example 1 below for Hi-Lo pricing, $z_{it}(j)$ has two components.} That is:
    \begin{equation}
        p_{it}(j)
        \text{ } = \text{ }
        \rho \left(
            z_{it}(j),  \text{ } e_{it}(j)
        \right)
    \label{eq_process_price}
    \end{equation}
    where $\rho(.)$ is a function that is strictly increasing in all its arguments, and it is known to the researcher. (A) The vector of persistent components $\mathbf{z}_{it} \equiv (z_{it}(j):j=1,2,...,J)$ follows a first order Markov process with transition density function $f_{z}(\mathbf{z}_{i,t+1}|\mathbf{z}_{it})$. (B) (Conditional independence of the transitory component) Conditional on $\mathbf{z}_{it}$, we have that $\mathbf{z}_{i,t+1}$ and $\mathbf{e}_{i,t+1}$ do not depend on $\mathbf{e}_{it}$ or on previous lags of this transitory component.  (C) The vectors $\mathbf{z}_{it}-\mathbf{z}_{i,t-1}$ and $\mathbf{e}_{it}-\mathbf{e}_{i,t-1}$ have supports that include a neighborhood around zero.} $\qquad \blacksquare$
\label{assumption_1}
\end{assumption}

\bigskip

Some examples for the $\rho(.)$ function are the linear function $p_{it}(j) = z_{it}(j) + e_{it}(j)$, or the linear in logs function, $p_{it}(j) = \exp\{z_{it}(j)  e_{it}(j)\}$. See also the Example below for the case of Hi-Lo pricing. The stochastic relationship between the prices of the different products is unrestricted.

\bigskip

\begin{figure}[ht]
\caption{Weekly time series of price of a laundry detergent product (Tide liquid 70oz) \label{fig: hilo_laundry}}
    \centering
    \includegraphics[width=14cm]{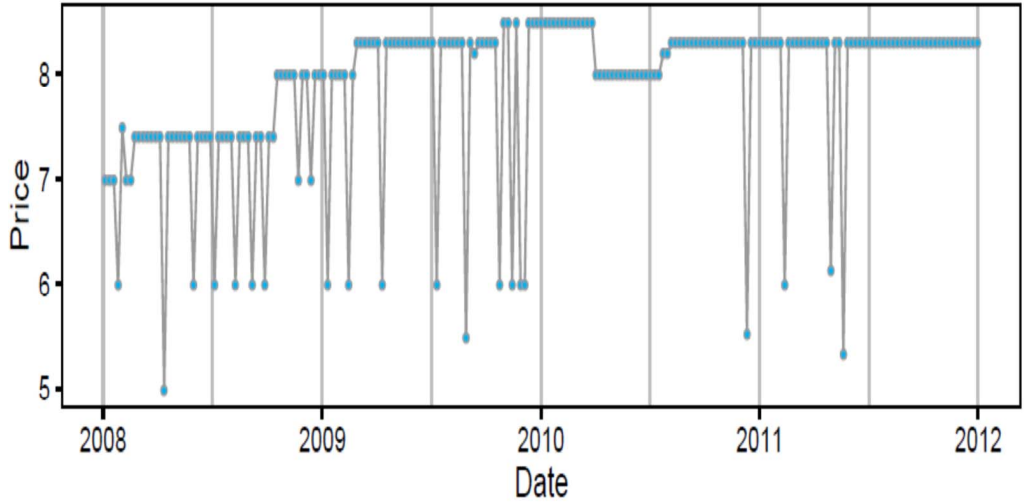}
\end{figure}

\noindent \textbf{EXAMPLE. Hi-Lo pricing.} For many supermarket products, the evolution of weekly prices is characterized by the alternation between 
a regular price and a promotion price. The (latent) regular and promotion prices remain constant for months, but the occurrence of a promotion is very uncertain and it lasts no more than one week. See, for instance, \citeauthor{hitsch_hortacsu_2019}
(\citeyear{hitsch_hortacsu_2019}) for recent and extensive evidence using Nielsen data. Figure \ref{fig: hilo_laundry} presents an example of the actual time series of prices for a laundry detergent product. \citeauthor{liu_balachander_2014}
(\citeyear{liu_balachander_2014}) show that this evolution of prices is not consistent with the assumption of first order Markov process that is common in many empirical applications of dynamic demand models. They find that relaxing this restriction has important implications to explain consumer behavior.

I present here a stochastic process for prices under Hi-Lo pricing that satisfies the general conditions described in \textcolor{blue}{Assumption} \ref{assumption_1} above. The persistent component $z_{it}(j)$ has two elements: the \textit{regular price}, $z_{it}^{reg}(j)$, and the \textit{promotion price}, $z_{it}^{pro}(j)$. By definition, $z_{it}^{reg}(j) > z_{it}^{pro}(j)$. These two (latent) variables have discrete support and follow a joint first order Markov process with substantial persistence.\footnote{More precisely, the whole vector $\mathbf{z}_{it}$ for all the $J$ products follows a first order Markov chain.} At any period $t$, the price of product $j$ can be either the regular price (i.e., $p_{it}(j) = z_{it}^{reg}(j)$), or the promotion price (i.e., $p_{it}(j) = z_{it}^{pro}(j)$). Let $e_{it}(j)$ be the dummy variable that indicates the existence of a promotion for product $j$ in market $i$ at period $t$. Therefore, in this example, the $\rho$ function has the following form:
\begin{equation}
    p_{it}(j)
    \text{ } = \text{ }
    \rho \left(
        z_{it}(j),  \text{ } e_{it}(j)
    \right)
    \text{ } = \text{ }
    \left( 1 - e_{it}(j) \right) \text{ } z_{it}^{reg}(j)
    \text{ } + \text{ }
    e_{it}(j) \text{ } z_{it}^{pro}(j)
\label{eq_hilo_pricing}
\end{equation}
Conditional on $\mathbf{z}_{it}$, the vector of promotion dummies $\mathbf{e}_{it}$ has a multinomial distribution that is independently distributed over time. 

\citeauthor{liu_balachander_2014}
(\citeyear{liu_balachander_2014}) consider a stochastic process for prices that satisfies equation \eqref{eq_hilo_pricing} but where the latent variables for regular and promotion prices are constant over time, and the sales promotion dummy follows a Proportional Hazard Model with duration dependence. Their stochastic process satisfies \textcolor{blue}{Assumption} \ref{assumption_1}(A) and \textcolor{blue}{Assumption} \ref{assumption_1}(B) as long as we include the variable \textit{“duration since last sales promotion”} as part of the persistent component, $\mathbf{z}_{it}$. However, it does not satisfy \textcolor{blue}{Assumption} \ref{assumption_1}(C) because the duration variable never remains constant over time, such that the distribution of $duration_{it}-duration_{i,t-1}$ does not have probability mass around zero. A condition that makes Liu-Balachander’s PHM consistent with \textcolor{blue}{Assumption} \ref{assumption_1} is that there exists a value of \textit{“duration since last sales promotion”}, say $D$, such that there is no duration dependence either for $duration_{it}<D$ or for $duration_{it}>D$. This restriction is testable. $\qquad \blacksquare$


This structure has an important implication on a consumer's dynamic decision model. The whole vector of prices $\mathbf{p}_{t}$ (both $\mathbf{z}_{t}$ and $\mathbf{e}_{t}$) affects a consumer's current utility, but the expected and discounted value of future utilities (the continuation value) depends on $\mathbf{z}_{t}$ but not on $\mathbf{e}_{t}$. This exclusion restriction plays an important role in the identification of parameter $\gamma$ in the Fixed Effects dynamic forward-looking model in this paper.

It is relevant to note that, given time series data on prices and a specification of the $\rho(.)$ function, it is possible to identify the parameters in the two stochastic processes that characterize the evolution of prices, and based on these parameters, it is possible to identify (i.e., to estimate consistently) the two components $\mathbf{z}_{it}$ and $\mathbf{e}_{it}$. In the Example of Hi-Lo pricing above, this identification is straightforward. For the rest of the paper, I assume that these two vectors, $\mathbf{z}_{it}$ and $\mathbf{e}_{it}$, are observable to the researcher.

\subsection{Consumer decision problem}

Every period $t$, the consumer observes the vectors of state variables $\mathbf{x}_{it}$, $\mathbf{z}_{it}$, $\mathbf{e}_{it}$, and $\boldsymbol{\varepsilon}_{it}$, and makes a purchasing decision $y_{it}$ to maximize her expected and discounted intertemporal utility $\mathbb{E}_{t}\left[
{\textstyle\sum\nolimits_{s=0}^{\infty}}
\text{ } \delta_{i}^{s} \text{ } U_{i,t+s} \right]$, where $\delta_{i} \in [0,1)$ is consumer $i$'s time discount factor.\footnote{Consumers know $\mathbf{z}_{it}$ and $\mathbf{e}_{it}$. For instance, in the Example for Hi-Lo pricing above, consumers can distinguish between regular and promotion prices.} This consumer's problem is a stationary \textit{Markov Decision Process (MDP)}, and Blackwell’s Theorem establishes that the value function and the optimal decision rule are time-invariant (\citeauthor{blackwell_1965}, \citeyear{blackwell_1965}).

The decision problem of consumer $i$ at period $t$ is:
\begin{equation}
    y_{it} \text{ } = \text{ }
    \argmax_{j \in \mathcal{Y}}
    \left\{ 
        \text{ }
        u_{\boldsymbol{\alpha}_{i}}(j,\mathbf{x}_{it},\mathbf{p}_{it})
        \text{ } + \text{ }
        \varepsilon_{it}(j)
        \text{ } + \text{ }
        v_{\boldsymbol{\alpha}_{i}}(f_{x}(j,\mathbf{x}_{it}),\mathbf{z}_{it})
        \text{ }
    \right\}
\label{eq_optimal_decision}
\end{equation}
where, as defined in equation \eqref{eq_transition_endogenous}, $f_{x}(j,\mathbf{x}_{it})$ represents the value of $\mathbf{x}_{i,t+1}$ given state $\mathbf{x}_{it}$ and decision $y_{it}=j$; and $v_{\boldsymbol{\alpha}_{i}}(f_{x}(j,\mathbf{x}_{it}),\mathbf{z}_{it})$ is the \textit{continuation value function}, i.e., the expected and discounted value of future utility given current state is $(\mathbf{x}_{it},\mathbf{z}_{it})$ and current choice $j$.\footnote{In the literature of structural dynamic discrete models, letter $v$ is commonly used to denote the \textit{choice-specific value function} or \textit{conditional-choice value function}. Instead, here I use letter $v$ to denote the continuation value, such that the \textit{choice-specific value function} is $u_{\boldsymbol{\alpha}_{i}}(j,\mathbf{x}_{it},\mathbf{p}_{it}) + v_{\boldsymbol{\alpha}_{i}}(f_{x}(j,\mathbf{x}_{it}),\mathbf{z}_{it})$.} 

A property of the model that plays a key role in the identification of parameter $\gamma$ is that the continuation value function $v_{\boldsymbol{\alpha}_{i}}(f_{x}(j,\mathbf{x}_{it}),\mathbf{z}_{it})$ does not depend on the transitory component of prices, $\mathbf{e}_{it}$. To see this, note that the continuation value is the expectation of next period value function, say $V_{\boldsymbol{\alpha}_{i}}(\mathbf{x}_{i,t+1},\mathbf{z}_{i,t+1},\mathbf{e}_{i,t+1})$, conditional on the current value of state variables at period $t$, $(\mathbf{x}_{it},\mathbf{z}_{it},\mathbf{e}_{it})$, and on current consumer choice, $y_{it}$. To obtain this expectation, one needs to integrate next period value function over the transition density function of the price-related state variables, say $f_{z,e}(\mathbf{z}_{i,t+1},\mathbf{e}_{i,t+1} |\mathbf{z}_{it},\mathbf{e}_{it})$. The conditional independence in
\textcolor{blue}{Assumption} \ref{assumption_1}(B) implies that this transition density does not depend on $\mathbf{e}_{it}$: that is, $f_{z,e}(\mathbf{z}_{i,t+1},\mathbf{e}_{i,t+1} |\mathbf{z}_{it},\mathbf{e}_{it}) = f_{z,e}(\mathbf{z}_{i,t+1},\mathbf{e}_{i,t+1} |\mathbf{z}_{it})$. The value of the transitory component of prices at period $t$ does not contain any information about prices at period $t+1$. This implies that the continuation value function does not depend on $\mathbf{e}_{it}$. Note that this property does not depend on any functional form assumption in the utility function, but on the conditional independence in 
\textcolor{blue}{Assumption} \ref{assumption_1}(B).

Define the \textit{conditional choice probability} (CCP) function as:
\begin{equation}
    \begin{array}[c]{l}
    P(j | \mathbf{x}_{it},\mathbf{z}_{it},\mathbf{e}_{it}, \boldsymbol{\alpha}_{i})
    \equiv 
    \\
    \\
    \qquad
    Pr\left( 
        j =
        \displaystyle \argmax_{k \in \mathcal{Y}}
        \bigg\{ 
            \text{ }
            u_{\boldsymbol{\alpha}_{i}}(k,\mathbf{x}_{it},\mathbf{p}_{it})
            + \varepsilon_{it}(k) +          
            v_{\boldsymbol{\alpha}_{i}}(f_{x}(k,\mathbf{x}_{it}),\mathbf{z}_{it})
        \text{ }
        \bigg\} 
        \text{ } \bigg|  \text{ }
        \mathbf{x}_{it},\mathbf{z}_{it}, \mathbf{e}_{it}, \boldsymbol{\alpha}_{i}
    \right)
    \end{array}
\label{eq_ccp_function}
\end{equation}
This CCP function is conditional on the observable state variables $(\mathbf{x}_{it},\mathbf{z}_{it},\mathbf{e}_{it})$ and on the unobservable fixed effects $\boldsymbol{\alpha}_{i}$. The type I Extreme Value distribution of the unobservables $\boldsymbol{\varepsilon}$ implies that the CCP function has the following form. For any $j \in \mathcal{Y}$:
\begin{equation}
    P(j | \mathbf{x}_{it},\mathbf{z}_{it},\mathbf{e}_{it}, \boldsymbol{\alpha}_{i}) =
    \displaystyle \frac{
        \exp \left\{ 
            u_{\boldsymbol{\alpha}_{i}}(j,\mathbf{x}_{it},\mathbf{p}_{it}) +
            v_{\boldsymbol{\alpha}_{i}}(f_{x}(j,\mathbf{x}_{it}),\mathbf{z}_{it})
        \right\}
    }
    {\sum_{k=0}^{J}
     \exp \left\{ 
        u_{\boldsymbol{\alpha}_{i}}(k,\mathbf{x}_{it},\mathbf{p}_{it}) +
        v_{\boldsymbol{\alpha}_{i}}(f_{x}(k,\mathbf{x}_{it}),\mathbf{z}_{it})
    \right\}
    }
\label{eq_logit_ccp_function}
\end{equation}
The logarithm of this CCP function is: 
\begin{equation}
    \log P \left(
        j | \mathbf{x}_{it},\mathbf{z}_{it},\mathbf{e}_{it}, \boldsymbol{\alpha}_{i}
    \right)
    \text{ } = \text{ }
    u_{\boldsymbol{\alpha}_{i}}(j,\mathbf{x}_{it},\mathbf{p}_{it}) 
    \text{ } + \text{ }
    v_{\boldsymbol{\alpha}_{i}}(f_{x}(j,\mathbf{x}_{it}),\mathbf{z}_{it})
    \text{ } - \text{ }
    \sigma_{\boldsymbol{\alpha}_{i}}(\mathbf{x}_{it},\mathbf{z}_{it},\mathbf{e}_{it})
\label{eq:log_CCP_function}
\end{equation}
where $\sigma_{\boldsymbol{\alpha}_{i}}(\mathbf{x}_{it},\mathbf{z}_{it},\mathbf{e}_{it})$ is the logarithm of the denominator in the logit CCP function:
\begin{equation}
    \sigma_{\boldsymbol{\alpha}_{i}}(\mathbf{x}_{it},\mathbf{z}_{it},\mathbf{e}_{it})
    \text{ } = \text{ }
    \log \left(
        {\displaystyle{\sum_{k=0}^{J}}}
        \exp \left\{
            u_{\boldsymbol{\alpha}_{i}}(k,\mathbf{x}_{it},\mathbf{p}_{it})
            \text{ } + \text{ }
            v_{\boldsymbol{\alpha}_{i}}(f_{x}(k,\mathbf{x}_{it}),\mathbf{z}_{it})
        \right\}
    \right)
\label{eq:bellman_equation}
\end{equation}


\section{Identification \label{sec_identification}}

The researcher observes a panel dataset of $N$ households over $T$ periods with information on households' purchasing decisions and prices. The time length of the panel, $T$, is short in the sense that it contains only a few purchases per household. The identification results in this paper consider that $T$ is fixed and -- as it is common in proofs of identification -- that $N$ is infinite such that we have an infinite population of households. Also, remember that variables $\mathbf{z}_{it}$ and $\mathbf{e}_{it}$ have been retrieved using data on prices, such that they are observable to the researcher. 

\textcolor{blue}{Assumption} \ref{assumption_2}, together with \textcolor{blue}{Assumption} \ref{assumption_1}, summarize the restrictions on the model for the identification of structural parameters $\widetilde{\beta}^{sc}$, $\beta^{dep}$, and $\gamma$.

\bigskip

\begin{assumption}
    \textit{(A) (i.i.d. Logit shocks) $\varepsilon_{it}(j)$ is mean-zerp $i.i.d.$ over $(i,t,j)$ with type I Extreme Value distribution, and is independent of $\boldsymbol{\alpha}_{i}$. (B) (Strict exogeneity of prices with respect to shocks $\boldsymbol{\varepsilon}_{it}$) For any two periods, $t$ and $s$, the variables $\varepsilon_{it}(j)$ and prices $\mathbf{p}_{is}$ are independently distributed.} $\qquad \blacksquare$
\label{assumption_2}
\end{assumption}

\bigskip

\textcolor{blue}{Assumption} \ref{assumption_2}(A) and \ref{assumption_2}(B) rule out product-time effects in consumer utility which are common to all consumers in the same local market and unobservable to the researcher. This restriction is very common in the literature of structural dynamic demand models, especially in applications using consumer level data.\footnote{For instance, \citeauthor{erdem_imai_2003} (\citeyear{erdem_imai_2003}), \citeauthor{hendel_nevo_ecma_2006} (\citeyear{hendel_nevo_ecma_2006}), \citeauthor{pavlidis_2017}, (\citeyear{pavlidis_2017}), \citeauthor{griffith_2018}, (\citeyear{griffith_2018}), or  \citeauthor{mysliwski_sanches_2020}, (\citeyear{mysliwski_sanches_2020}), among many other papers in this literature, assume that prices are not correlated with transitory shocks. In contrast, some dynamic demand models using market level data, such as \citeauthor{gowrisankaran_rysman_2012} (\citeyear{gowrisankaran_rysman_2012}), allow for unobserved market-level demand shocks.} It is important to note that, until very recently, there was not any point identification result for Fixed Effects dynamic logit models with time dummies (i.e., time fixed effects) using short panels.\footnote{The absence of this identification result is mentioned in \citeauthor{honore_kyriazidou_2000} (\citeyear{honore_kyriazidou_2000}), \citeauthor{hahn_2001} (\citeyear{hahn_2001}), or \citeauthor{honore_tamer_2006} (\citeyear{honore_tamer_2006}),  among many others. Existing fixed effects methods, such as \citeauthor{honore_kyriazidou_2000} (\citeyear{honore_kyriazidou_2000}), are based on matching values of exogenous explanatory variables over several time periods. This matching is not possible when the set of explanatory variables includes time dummies, or even a linear time trend. A recent paper by \citeauthor{kitazawa_2022} (\citeyear{kitazawa_2022}) shows identification of a Fixed Effects dynamic binary logit model with time dummies and $T=4$. His identification approach uses moment conditions based on an ingenious transformation of the model. Nevertheless, it seems difficult to extend his result to a model where agents are forward-looking because time effects appear also in the agent's continuation value function together with the individual fixed effect.} This may seem a strong assumption in an econometric demand model, and at odds with the literature on estimation of static demand models using the BLP framework (\citeauthor{berry_levinshon_1995}, \citeyear{berry_levinshon_1995}). However, as I explain in the next two paragraphs, the FE approach in this paper controls for two important sources of endogeneity in prices.

First, note that the vector of fixed effects $\boldsymbol{\alpha}_{i} = (\alpha_{i}(1), \alpha_{i}(2), ..., \alpha_{i}(J))$ includes product fixed effects: in fact, it accounts for any interaction of product effects and consumer effects. Therefore, these incidental parameters account for time-invariant differences in product quality that can be correlated with the cross-sectional variation in prices across products. That is, the model accounts -- in a very general way -- for endogeneity of price levels. This means that the potential concern with \textcolor{blue}{Assumption} \ref{assumption_2}(B) is because of potential endogeneity of time-variation in prices.

Second, all the identification results in this paper control for variation over time in the persistent component of prices, $\mathbf{z}_{it}$ and exploit only variation in the transitory component, $\mathbf{e}_{it}$. Therefore, the method controls for endogeneity of \textit{regular prices}.

Importantly, we consider a fixed effects (FE) model, in the sense that both the unconditional distribution 
$F_{\boldsymbol{\alpha}}(\boldsymbol{\alpha}_{i})$ and the
sequence of conditional distributions $F_{\boldsymbol{\alpha}|\mathbf{x},t}(\boldsymbol{\alpha}_{i}|\ell_{it},d_{it},\mathbf{z}_{it},\mathbf{e}_{it})$ for $t=1,2, ...,T$ are completely unrestricted functions.

\subsection{Sufficient statistics -- Conditional likelihood approach}

Let $\boldsymbol{\theta}$ be the vector of structural parameters:
\begin{equation}
    \boldsymbol{\theta}
    \text{ } = \text{ }
    \left(
        \gamma, \text{ } 
        \widetilde{\beta}^{sc}(k,j), \text{ } 
        \beta^{dep}(j): k,j \in \{1,2,...,J\} 
        \text{ with }  k>j
    \right)^{\prime}
\label{eq:definition_theta}
\end{equation}
Note that vector $\boldsymbol{\theta}$ has finite dimension: more precisely, it has $1+J+\frac{J(J-1)}{2}$ elements. To establish the identification of $\boldsymbol{\theta}$, I follow \citeauthor{aguirregabiria_gu_2021} (\citeyear{aguirregabiria_gu_2021}), who consider a \textit{sufficient statistic - conditional likelihood} approach for dynamic nonlinear panel data models in the spirit of \citeauthor{cox_1958} (\citeyear{cox_1958}), \citeauthor{rasch_1960} (\citeyear{rasch_1960}), \citeauthor{chamberlain_1985}, (\citeyear{chamberlain_1985}), or \citeauthor{honore_kyriazidou_2000}, (\citeyear{honore_kyriazidou_2000}).\footnote{For the autoregressive binary logit model with fixed effects, the derivation of sufficient statistics and the identification of the autoregressive parameter are already established in \citeauthor{cox_1958} (\citeyear{cox_1958}). See page 227 in that paper.} I start describing some general features of this approach in the context of the dynamic demand model. For notational simplicity, I omit the vector of observable consumer characteristics, $\mathbf{w}_{i}$, for the rest of the paper. 

Let $\mathbf{y}_{i} \equiv$ $(\ell_{i1}, d_{i1}, y_{i1},y_{i2},...,y_{iT})$ be the vector with consumer $i$'s choice history, including the initial condition $(\ell_{i1}, d_{i1})$, and let $\widetilde{\mathbf{z}}_{i} \equiv (\mathbf{z}_{i1}, \mathbf{z}_{i2}, ..., \mathbf{z}_{iT})$ and $\widetilde{\mathbf{e}}_{i} \equiv (\mathbf{e}_{i1}, \mathbf{e}_{i2}, ..., \mathbf{e}_{iT})$ be the vectors with the histories of the two components in prices. The probability of a choice history $\mathbf{y}_{i}$ conditional on the history of prices and parameters $\boldsymbol{\theta}$ and $\boldsymbol{\alpha}_{i}$ is:
\begin{equation}
    \mathbb{P}
    \left(
        \mathbf{y}_{i}|\widetilde{\mathbf{z}}_{i},
        \widetilde{\mathbf{e}}_{i},\boldsymbol{\alpha}_{i},
        \boldsymbol{\theta}
    \right) =  
    p^{\ast}_{1}(\ell_{i1}, d_{i1} | \boldsymbol{\alpha}_{i}) 
    \displaystyle \prod_{t=1}^{T} 
    \displaystyle \frac{
        \exp \left\{ 
            u_{\boldsymbol{\alpha}_{i}}(y_{it},\mathbf{x}_{it},\mathbf{p}_{it}) +
            v_{\boldsymbol{\alpha}_{i}}(f_{x}(y_{it},\mathbf{x}_{it}),\mathbf{z}_{it})
        \right\}
    }
    {\sum_{j=0}^{J}
     \exp \left\{ 
        u_{\boldsymbol{\alpha}_{i}}(j,\mathbf{x}_{it},\mathbf{p}_{it}) +
        v_{\boldsymbol{\alpha}_{i}}(f_{x}(j,\mathbf{x}_{it}),\mathbf{z}_{it})
    \right\}
    }
\label{eq_prob_choice_history}
\end{equation}
where $p^{\ast}_{1}(.)$ is the probability of the state variables at period $t=1$. Given definition of $\sigma_{\boldsymbol{\alpha}_{i}}(\mathbf{x}_{it},\mathbf{z}_{it},\mathbf{e}_{it})$ -- in equation \eqref{eq:bellman_equation} above -- as the logarithm of the denominator in the logit CCP function, we have that the log-probability of choice history $\mathbf{y}_{i}$ is:
\begin{equation}
\begin{array}[c]{rcl}
    \log \mathbb{P}
    \left(
        \mathbf{y}_{i}|\widetilde{\mathbf{z}}_{i},
        \widetilde{\mathbf{e}}_{i},\boldsymbol{\alpha}_{i},
        \boldsymbol{\theta}
    \right) 
    & = &
    \log p^{\ast}_{1}(\ell_{i1}, d_{i1} | \boldsymbol{\alpha}_{i})
    \\
    & + &
    \displaystyle \sum_{t=1}^{T}
    \bigg[
        u_{\boldsymbol{\alpha}_{i}}(y_{it},\mathbf{x}_{it},\mathbf{p}_{it}) +     
        v_{\boldsymbol{\alpha}_{i}}(f_{x}(y_{it},\mathbf{x}_{it}),\mathbf{z}_{it}) - \sigma_{\boldsymbol{\alpha}_{i}}\left(\mathbf{x}_{it},\mathbf{z}_{it},\mathbf{e}_{it}\right)
    \bigg]
    \end{array}
\label{eq:log_prob_choice_history}
\end{equation}
Taking into account the structure of the utility function and the transition rules of the state variables, we can develop further this expression to obtain:
\begin{equation}
\begin{array}[c]{rcl}
    \log \mathbb{P}
    \left(
        \mathbf{y}_{i}|\widetilde{\mathbf{z}}_{i},
        \widetilde{\mathbf{e}}_{i},\boldsymbol{\alpha}_{i},
        \boldsymbol{\theta}
    \right) 
    & = &
    \log p^{\ast}_{1}(\ell_{i1}, d_{i1} | \boldsymbol{\alpha}_{i})
    \\
    & + &
    \displaystyle \sum_{t=1}^{T} 1\{ y_{it}=0 \}
    \bigg[
        \alpha_{i}(\ell_{it}) + \gamma  \text{ } h_{it}(0) - \beta^{dep}(\ell_{it})d_{it} 
    \bigg]
    \\
    & + &
    \displaystyle \sum_{t=1}^{T} 1\{ y_{it}=0 \}
    \bigg[
        v_{\boldsymbol{\alpha}_{i}}(\ell_{it},d_{it}+1,\mathbf{z}_{it}) 
        - \sigma_{\boldsymbol{\alpha}_{i}}(\ell_{it},d_{it},\mathbf{z}_{it},\mathbf{e}_{it}) 
    \bigg]
    \\
    & + &
    \displaystyle \sum_{t=1}^{T} \sum_{j=1}^{J} 1\{ y_{it}=j \}
    \bigg[
        \alpha_{i}(j) + \gamma  \text{ } h_{it}(j) - \beta^{sc}(\ell_{it},j) 
    \bigg]
    \\
    & + &
    \displaystyle \sum_{t=1}^{T} \sum_{j=1}^{J} 1\{ y_{it}=j \}
    \bigg[
        v_{\boldsymbol{\alpha}_{i}}(j,1,\mathbf{z}_{it}) 
        - \sigma_{\boldsymbol{\alpha}_{i}}(\ell_{it},d_{it},\mathbf{z}_{it},\mathbf{e}_{it}) 
    \bigg]
    \end{array}
\label{eq:log_prob_choice_history_further}
\end{equation}
where $1\{.\}$ is the indicator function.

Based on equation \eqref{eq:log_prob_choice_history_further}, the log-probability of a choice history has the following structure:
\begin{equation}
    \begin{array}[c]{lcr}
        \log \mathbb{P}
        \left(
            \mathbf{y}_{i}|\widetilde{\mathbf{z}}_{i},
            \widetilde{\mathbf{e}}_{i},\boldsymbol{\alpha}_{i},
            \boldsymbol{\theta}
        \right) 
    & = &  
    \mathbf{s}(\mathbf{y}_{i},\widetilde{\mathbf{z}}_{i},
    \widetilde{\mathbf{e}}_{i})^{\prime} \text{ } \mathbf{g}(\boldsymbol{\alpha}_{i}) + \mathbf{c}(\mathbf{y}_{i},\widetilde{\mathbf{z}}_{i},
    \widetilde{\mathbf{e}}_{i})^{\prime} \text{ }
    \boldsymbol{\theta}
    \end{array}
\label{prob_choice_history}
\end{equation}
where $\mathbf{g}(\boldsymbol{\alpha}_{i})$ is a vector of functions of the incidental parameters $\boldsymbol{\alpha}_{i}$, and $\mathbf{s}_{i} \equiv \mathbf{s}(\mathbf{y}_{i},\widetilde{\mathbf{z}}_{i},\widetilde{\mathbf{e}}_{i})$ and $\mathbf{c}_{i} \equiv \mathbf{c}(\mathbf{y}_{i},\widetilde{\mathbf{z}}_{i},\widetilde{\mathbf{e}}_{i})$ are vectors of statistics, i.e., functions of $(\mathbf{y}_{i},\widetilde{\mathbf{z}}_{i},\widetilde{\mathbf{e}}_{i})$. 
Equation \eqref{eq:log_prob_choice_history_further} provides the elements in each of these vectors. That is:
\begin{equation}
    \begin{array}[c]{lcl}
        \mathbf{g}(\boldsymbol{\alpha}_{i})         
        & = &  
        \biggl( 
            \log p^{\ast}_{1}(j, d | \boldsymbol{\alpha}_{i}), \text{ }
            \alpha_{i}(j), \text{ }
            v_{\boldsymbol{\alpha}_{i}}(j,d,\mathbf{z}), \text{ } 
            \sigma_{\boldsymbol{\alpha}_{i}}(j,d,\mathbf{z},\mathbf{e}):
        \\
        &  &  
            \qquad
            j=1,2,...,J; \text{ } d=1,2,...,D_{T}; \text{ }
            (\mathbf{z},\mathbf{e}) \in \mathcal{Z}_{T} \times \mathcal{E}_{T}
        \biggr)
    \end{array}
\label{eq;vector_g}
\end{equation}
where $D_{T}$ is the maximum duration observed in the data, and $\mathcal{Z}_{T} \times \mathcal{E}_{T}$ are the sets of values of $\mathbf{z}_{it}$ and $\mathbf{e}_{it}$, respectively, observed in the data. The following equation presents each element in the vector of statistics $\mathbf{s}_{i}$, with its corresponding element in vector  $\mathbf{g}(\boldsymbol{\alpha}_{i})$.
\begin{equation}
    \mathbf{s}_{i} =
    \displaystyle \sum_{t=1}^{T}
    \left[
    \begin{array}[c]{l}
        1\{(\ell_{i1},d_{i1})=(j,d)\}, \\
        1\{(y_{it},\ell_{it})=(0,j)\} + 1\{y_{it}=j\}, \\
        1\{(y_{it},\mathbf{z}_{it})=(j,\mathbf{z})\}, \\
        1\{(y_{it},\ell_{it},d_{it},\mathbf{z}_{it})=(0,j,d-1,\mathbf{z})\}, \\
        1\{(\ell_{it},d_{it},\mathbf{z}_{it},\mathbf{e}_{it})=(j,d,\mathbf{z},\mathbf{e})\}
    \end{array}
    \right]
    \rightarrow
    \left[
    \begin{array}[c]{l}
        \log p^{\ast}_{1}(j, d | \boldsymbol{\alpha}_{i}), \\
        \alpha_{i}(j), \\
        v_{\boldsymbol{\alpha}_{i}}(j,1,\mathbf{z}), \\
        v_{\boldsymbol{\alpha}_{i}}(j,d,\mathbf{z}) 
        \text{ } for \text{ } d \geq 2,  \\
        \sigma_{\boldsymbol{\alpha}_{i}}(j,d,\mathbf{z},\mathbf{e})
    \end{array}
    \right]
\label{eq;vector_s}
\end{equation}
Similarly, the following equation presents each element in the vector of statistics $\mathbf{c}_{i}$, with its corresponding element in the vector  of structural parameters $\boldsymbol{\theta}$.
\begin{equation}
    \mathbf{c}_{i} =
    \displaystyle \sum_{t=1}^{T}
    \left[
    \begin{array}[c]{l}
        1\{(y_{it},\ell_{it})=(0,j)\}h_{it}(0) + 1\{y_{it}=j\}h_{it}(j), \\
        -1\{(y_{it},\ell_{it})=(j,k)\}, \\
        -1\{(y_{it},\ell_{it})=(0,j)\}d_{it}
    \end{array}
    \right]
    \rightarrow
    \left[
    \begin{array}[c]{l}
        \gamma, \\
        \widetilde{\beta}^{sc}(k,j), \\
        \beta^{dep}(j)
    \end{array}
    \right]
\label{eq;vector_c}
\end{equation}

The structure in equation (\ref{prob_choice_history}) has key implications for the identification of $\boldsymbol{\theta}$. Suppose that, for every parameter in the vector $\boldsymbol{\theta}$, say $\theta_{k}$ (the k-th element of vector $\boldsymbol{\theta}$), there exist two choice histories, say $A$ and $B$, such that $\mathbf{s}(A) = \mathbf{s}(B)$ and $\mathbf{c}(A) - \mathbf{c}(B)$ is a vector where all the elements are zero except element $k$ that is one. Under these conditions, equation \eqref{prob_choice_history} implies that:
\begin{equation}
    \theta_{k} \text{ } = \text{ }
    \log \mathbb{P}(A) - \log \mathbb{P}(B),
\end{equation}
which shows that parameter $\theta_{k}$ is identified from the log odds ratio of histories $A$ and $B$.\footnote{Note that, in this class of models, every choice history has a strictly positive probability such that $\log \mathbb{P}(A)$ and $\log \mathbb{P}(B)$ are finite real numbers.} This intuitive description of the identification of parameters in FE discrete choice models has been used by \citeauthor{chamberlain_1985} (\citeyear{chamberlain_1985}) and \citeauthor{honore_kyriazidou_2000} (\citeyear{honore_kyriazidou_2000}), among others. In this paper, I follow this approach.

More generally, equation (\ref{prob_choice_history}) implies that $\mathbf{s}_{i}$ is a \textit{sufficient statistic} for $\boldsymbol{\alpha}_{i}$. That is, the probability $\mathbb{P}( \mathbf{y}_{i} | \widetilde{\mathbf{z}}_{i},\widetilde{\mathbf{e}}_{i}, \boldsymbol{\alpha}_{i}, \boldsymbol{\theta}, \mathbf{s}_{i})$ does not depend on $\boldsymbol{\alpha}_{i}$. To see this, note that $\mathbf{s}_{i}$ is a deterministic function of $\mathbf{y}_{i}$ such that $\mathbb{P}( \mathbf{y}_{i}, \mathbf{s}_{i}) =$ $\mathbb{P}( \mathbf{y}_{i})$. Then, applying equation \eqref{prob_choice_history}, we have:
\begin{equation}
\begin{array}
[c]{ccl}
    \mathbb{P}( \mathbf{y}_{i} | \widetilde{\mathbf{z}}_{i},\widetilde{\mathbf{e}}_{i}, \boldsymbol{\alpha}_{i}, \boldsymbol{\theta}, \mathbf{s}_{i}) 
    & = & 
    \dfrac{\mathbb{P}( \mathbf{y}_{i} | \widetilde{\mathbf{z}}_{i},\widetilde{\mathbf{e}}_{i}, \boldsymbol{\alpha}_{i}, \boldsymbol{\theta})}
    {\mathbb{P}( \mathbf{s}_{i} | \widetilde{\mathbf{z}}_{i},\widetilde{\mathbf{e}}_{i}, \boldsymbol{\alpha}_{i}, \boldsymbol{\theta}) }
    =
    \dfrac{
    \exp\left\{  
        \mathbf{s}_{i}^{\prime} \text{ } \mathbf{g}(\boldsymbol{\alpha}_{i}) + \mathbf{c}_{i}^{\prime} \text{ } \boldsymbol{\theta}
    \right\}  }
    {\sum_{\mathbf{y}: \text{ } 
    \mathbf{s}(\mathbf{y}) = \mathbf{s}_{i}}
    \exp\left\{  
        \mathbf{s}_{i}^{\prime} \text{ } \mathbf{g}(\boldsymbol{\alpha}_{i}) + \mathbf{c}(\mathbf{y})^{\prime} \text{ }
        \boldsymbol{\theta}
    \right\}  } \\
    &  & \\
    & = & 
    \dfrac{
    \exp\left\{  
         \mathbf{c}_{i}^{\prime} \text{ } \boldsymbol{\theta}
    \right\}  }
    {\sum_{\mathbf{y}: \text{ } 
    \mathbf{s}(\mathbf{y}) = \mathbf{s}_{i}}
    \exp\left\{  
         \mathbf{c}(\mathbf{y})^{\prime} \text{ }
        \boldsymbol{\theta}
    \right\}  }
    =
    \mathbb{P}( \mathbf{y}_{i} | \mathbf{s}_{i}, \mathbf{c}_{i}, \boldsymbol{\theta}) 
\end{array}
\label{sufficiency}
\end{equation}
where $\sum_{\mathbf{y:}\text{ }\mathbf{s(y)=s}_{i}}$ is the sum over all the possible choice histories $\mathbf{y}$ with $\mathbf{s(y)}$ equal to $\mathbf{s}_{i}$. This sufficiency result comes from usual properties of the exponential family, to which the Type I Extreme Value distribution belongs.

Second, using the expression at the bottom line of equation (\ref{sufficiency}), we have the following conditional log-likelihood function (at the population level):
\begin{equation}
    \mathbb{E} \left[
        \log  \mathbb{P}( \mathbf{y}_{i} | \widetilde{\mathbf{z}}_{i},\widetilde{\mathbf{e}}_{i}, \boldsymbol{\theta}, \mathbf{s}_{i})
    \right]
    \text{ } = \text{ }
    \mathbb{E} \left[
        \mathbf{c}_{i}^{\prime} \text{ } \boldsymbol{\theta} 
        \text{ } - \text{ }
        \log \left(
            \sum_{\mathbf{y}: \text{ } 
            \mathbf{s}(\mathbf{y}) = \mathbf{s}_{i}} 
            \exp\left\{          
                \mathbf{c}(\mathbf{y})^{\prime} \text{ }
                \boldsymbol{\theta}
            \right\} 
        \right)
    \right]
\end{equation}
The first order conditions for the maximization of this likelihood function with respect to $\boldsymbol{\theta}$ imply the following moment conditions (i.e., likelihood equations):
\begin{equation}
    \boldsymbol{m} \left( \boldsymbol{\theta} \right)
    \text{ } \equiv \text{ }
    \mathbb{E} \left[
        \mathbf{c}_{i}
        \text{ } - \text{ }
        \sum_{\mathbf{y}: \text{ } 
        \mathbf{s}(\mathbf{y}) = \mathbf{s}_{i}} 
        \mathbf{c}(\mathbf{y}) \text{ }
        \mathbb{P}( \mathbf{y} | \widetilde{\mathbf{z}}_{i},\widetilde{\mathbf{e}}_{i}, \boldsymbol{\theta}, \mathbf{s}_{i}) 
    \right]
    \text{ } = \text{ } \mathbf{0}
\label{eq:general_moment_conditions}
\end{equation}
The Jacobian matrix for this vector of moment conditions -- or equivalently, the Hessian of the log-likelihood function, or the negative of Fisher's information matrix -- is:
\begin{equation}
    \frac{\partial \boldsymbol{m} \left( \boldsymbol{\theta} \right)}
    {\partial \boldsymbol{\theta}^{\prime}}
    \text{ } = \text{ }
    - \mathbb{E} \left(
        \left[
            \mathbf{c}_{i} - \mathbb{E}(\mathbf{c}_{i}|\mathbf{s}_{i})
        \right] 
        \text{ } 
        \left[
            \mathbf{c}_{i} - \mathbb{E}(\mathbf{c}_{i}|\mathbf{s}_{i})
        \right]^{\prime}
    \right)
\label{eq:jacobian_moments}
\end{equation}
This is the negative of a variance-covariance matrix, and therefore, it is negative semidefinite for any value of $\boldsymbol{\theta}$. In other words, the likelihood function is globally concave in $\boldsymbol{\theta}$. Furthermore, based on these moment conditions, a necessary and sufficient condition for (point) identification of $\boldsymbol{\theta}$ is that this Jacobian matrix $\partial \boldsymbol{m}(\boldsymbol{\theta})/\partial\boldsymbol{\theta}^{\prime}$ is nonsingular (i.e., rank identification condition). The right-hand-side in equation \eqref{eq:jacobian_moments} is a nonsingular matrix if there is
no perfect collinearity between the residuals of vector $\mathbf{c}_{i}$ once we remove its mean dependence on $\mathbf{s}_{i}$.

\bigskip

It is important to note that for some FE nonlinear panel data models the identification conditions based on this conditional likelihood approach are sufficient but not necessary. \citeauthor{bonhomme_2012} (\citeyear{bonhomme_2012}) provides a systematic approach (the \textit{functional differencing} method) to construct moment restrictions for a general class of FE models. Recently, 
\citeauthor{kitazawa_2013} (\citeyear{kitazawa_2013}, \citeyear{kitazawa_2022}), \citeauthor{honore_weidner_2020} (\citeyear{honore_weidner_2020}), \citeauthor{dobronyi_gu_2021} (\citeyear{dobronyi_gu_2021}), and 
\citeauthor{honore_muris_2021} (\citeyear{honore_muris_2021}) have used Bonhomme's \textit{functional differencing} approach to establish new point identification results of parameters in FE models where the conditional likelihood approach, at least apparently, does not provide identification.

For the rest of this subsection, I present results on the identification of structural parameters in two versions of the demand model: with and without duration dependence. Here, for simplicity, I do not present the expression for the statistics $\mathbf{s}_{i}$ and $\mathbf{c}_{i}$ and instead provide examples of pairs of histories $A$ and $B$ that identify the different parameters in $\boldsymbol{\theta}$. 

\subsection{Dynamic demand without duration dependence}

Suppose that there is no depreciation or depletion, so that $\beta^{dep}(j) = 0$ for every product $j$, but there is habit formation or/and brand switching costs, so that $\beta^{sc}(k,j) \neq 0$ for some pair of products $(k,j)$. This is the demand model in empirical applications like \citeauthor{roy_chintagunta_1996} (\citeyear{roy_chintagunta_1996}), \citeauthor{keane_1997} (\citeyear{keane_1997}), \citeauthor{osborne_2011} (\citeyear{osborne_2011}), or  \citeauthor{mysliwski_sanches_2020}, (\citeyear{mysliwski_sanches_2020}), among others. 

In this restricted version of the model, $\ell_{it}$ is the only state variable whose evolution depends on the consumer's choices. Its transition function is:
\begin{equation}
    \ell_{i,t+1}
    \text{ } = \text{ }
    f_{x}(y_{it},\ell_{it})
    \text{ } = \text{ }
    \left\{
    \begin{array}
    [c]{rcl}
    \ell_{it} & if & y_{it}=0
    \\
    j & if & y_{it}=j>0  
    \end{array}
    \right. 
\label{eq_transition_only_last}
\end{equation}
Then, combining equation \eqref{eq:log_CCP_function} with this transition rule, we have the following expression for the log-CCP function that, for simplicity, I represent using the shorter notation $\log P_{it}$ instead of $\log P \left(y_{it} | \mathbf{x}_{it},\mathbf{z}_{it},\mathbf{e}_{it}, \boldsymbol{\alpha}_{i}\right)$: 
\begin{equation}
    \log P_{it} =
    \left\{
    \begin{array}
    [c]{lcl}
    \alpha_{i}(\ell_{it}) + 
    v_{\boldsymbol{\alpha}_{i}}(\ell_{it},\mathbf{z}_{it}) -
    \sigma_{\boldsymbol{\alpha}_{i}}\left(\ell_{it},\mathbf{z}_{it},\mathbf{e}_{it}\right)
    & if & y_{it}=0
    \\
    \alpha_{i}(j) + \gamma \text{ } h_{it}(j) - \beta^{sc}(\ell_{it},j) +
    v_{\boldsymbol{\alpha}_{i}}(j,\mathbf{z}_{it}) -
    \sigma_{\boldsymbol{\alpha}_{i}}\left(\ell_{it},\mathbf{z}_{it},\mathbf{e}_{it}\right)
    & if & y_{it}=j>0
    \end{array}
    \right. 
\label{eq_utility_branded_product}
\end{equation}
where, for notational simplicity, I use $h_{it}(j)$ to represent $h\left(\mu_{i} - p_{it}(j) \right)$.

Suppose that $T=4$. Let $k$ and $j$ be two different products, and consider the following pair of choice histories:
\begin{equation}
    A 
    \text{ } = \text{ }
    \left(
        k, \text{ } 
        j, \text{ }
        k, \text{ }
        j
    \right)
    \quad ; \quad
        B
    \text{ } = \text{ }
    \left(
        k, \text{ } 
        k, \text{ }
        j, \text{ }
        j
    \right) 
\label{eq_example_1_histories_nozeros}
\end{equation}
Taking into account the structure of the log-CCP function in equation \eqref{eq_utility_branded_product}, we have the following expression for the log-probabilities of the choice histories:
\begin{equation}
    \begin{array}
    [c]{ccl}
    \log \mathbb{P}(A)
    & = &
    \log p^{\ast}_{1}(k,\boldsymbol{\alpha}_{i}) +
    \alpha_{i}(j) + \alpha_{i}(k) + \alpha_{i}(j) \\
    & + &
    v_{\boldsymbol{\alpha}_{i}}(j,\mathbf{z}_{2}) +
    v_{\boldsymbol{\alpha}_{i}}(k,\mathbf{z}_{3}) +
    v_{\boldsymbol{\alpha}_{i}}(j,\mathbf{z}_{4}) \\
    & - &
    \sigma_{\boldsymbol{\alpha}_{i}}(k,\mathbf{z}_{2},\mathbf{e}_{2}) -
    \sigma_{\boldsymbol{\alpha}_{i}}(j,\mathbf{z}_{3},\mathbf{e}_{3}) -
    \sigma_{\boldsymbol{\alpha}_{i}}(k,\mathbf{z}_{4},\mathbf{e}_{4}) \\
    & - &
    \beta^{sc}(k,j) - \beta^{sc}(j,k) - \beta^{sc}(k,j) + 
    \gamma \text{ } 
    \left( h_{2}(j) + h_{3}(k) + h_{4}(j) \right),
    \end{array}
\label{eq_logPA_noduration}
\end{equation}
and
\begin{equation}
    \begin{array}
    [c]{ccl}
    \log \mathbb{P}(B)
    & = &
    \log p^{\ast}_{1}(k,\boldsymbol{\alpha}_{i}) +
    \alpha_{i}(k) + \alpha_{i}(j) + \alpha_{i}(j) \\
    & + &
    v_{\boldsymbol{\alpha}_{i}}(k,\mathbf{z}_{2}) +
    v_{\boldsymbol{\alpha}_{i}}(j,\mathbf{z}_{3}) +
    v_{\boldsymbol{\alpha}_{i}}(j,\mathbf{z}_{4}) \\
    & - &
    \sigma_{\boldsymbol{\alpha}_{i}}(k,\mathbf{z}_{2},\mathbf{e}_{2}) -
    \sigma_{\boldsymbol{\alpha}_{i}}(k,\mathbf{z}_{3},\mathbf{e}_{3}) -
    \sigma_{\boldsymbol{\alpha}_{i}}(j,\mathbf{z}_{4},\mathbf{e}_{4}) \\
    & - &
    \beta^{sc}(k,j) + 
    \gamma \text{ } 
    \left( h_{2}(k) + h_{3}(j) + h_{4}(j) \right),
    \end{array}
\label{eq_logPB_noduration}
\end{equation}
Therefore, the difference between log-probabilities of the two histories is:
\begin{equation}
    \begin{array}
    [c]{ccl}
    \log \mathbb{P}(A) - \log \mathbb{P}(B)
    & = &
    v_{\boldsymbol{\alpha}_{i}}(j,\mathbf{z}_{2}) +
    v_{\boldsymbol{\alpha}_{i}}(k,\mathbf{z}_{3}) -
    v_{\boldsymbol{\alpha}_{i}}(k,\mathbf{z}_{2}) -
    v_{\boldsymbol{\alpha}_{i}}(j,\mathbf{z}_{3}) \\
    & - &
    \sigma_{\boldsymbol{\alpha}_{i}}(j,\mathbf{z}_{3},\mathbf{e}_{3}) -
    \sigma_{\boldsymbol{\alpha}_{i}}(k,\mathbf{z}_{4},\mathbf{e}_{4}) + 
    \sigma_{\boldsymbol{\alpha}_{i}}(k,\mathbf{z}_{3},\mathbf{e}_{3}) +
    \sigma_{\boldsymbol{\alpha}_{i}}(j,\mathbf{z}_{4},\mathbf{e}_{4}) \\
    & - &
    \widetilde{\beta}^{sc}(k,j) + 
    \gamma \text{ } 
    \left( h_{2}(j) - h_{3}(j) - h_{2}(k) + h_{3}(k) \right),
    \end{array}
\label{eq_logP_difference_noduration}
\end{equation}

Given equation \eqref{eq_logP_difference_noduration}, we can establish the identification of parameters $\widetilde{\beta}^{sc}(k,j)$ and $\gamma$. To control for the unobserved heterogeneity in the functions $v(.,\boldsymbol{\alpha}_{i})$ and $\sigma(.,\boldsymbol{\alpha}_{i})$, we need to impose the restrictions $\mathbf{z}_{2} = \mathbf{z}_{3} = \mathbf{z}_{4}$ and $\mathbf{e}_{3} = \mathbf{e}_{4}$. Given these restrictions, we have that:
\begin{equation}
    \log \mathbb{P}(A) - \log \mathbb{P}(B) =
    - \widetilde{\beta}^{sc}(k,j) + 
    \gamma \text{ } 
    \left( h_{2}(j) - h_{3}(j) - h_{2}(k) + h_{3}(k) \right),
\label{eq_logP_difference_final_noduration}
\end{equation}
Identification of parameters $\widetilde{\beta}^{sc}(k,j)$ and $\gamma$ follows from equation \eqref{eq_logP_difference_final_noduration}. When $p_{2}(j) - p_{3}(j)=0$ and $p_{2}(k) - p_{3}(k)=0$ -- implying that $h_{2}(j) - h_{3}(j) - h_{2}(k) + h_{3}(k) = 0$ -- we have identification of the switching cost parameter, i.e., $\widetilde{\beta}^{sc}(k,j) = - \log \mathbb{P}(A) + \log \mathbb{P}(B)$. Given the identified switching cost and condition $h_{2}(j) - h_{3}(j) - h_{2}(k) + h_{3}(k) \neq 0$, we have identification of the price-sensitivity parameter $\gamma$. Remember that equation \eqref{eq_logP_difference_final_noduration} is derived under the condition that the persistent component of prices is constant between periods 2 and 4, and the transitory component is constant between periods 3 and 4. Under these restrictions, we still can have $h_{2}(j) - h_{3}(j) - h_{2}(k) + h_{3}(k) \neq 0$ by using variation between periods 2 and 3 in the transitory components of prices of products $j$ or $k$. For instance, in the Example of Hi-Lo pricing, we can have a sales promotion for product $j$ but not for $k$, or vice-versa. 

This identification result applies to any functional form of $h(.)$, linear or nonlinear, as long as the researcher knows this function and observes consumer disposable income. Similarly, this identification result applies to any form of function $\rho(.)$ in the stochastic process of prices, given that this function is known to the researcher and the transitory component of prices satisfies the conditional independence in \textcolor{blue}{Assumption} \ref{assumption_1}(B).\footnote{On a minor technical point, remember that both $h(.)$ and $\rho(.)$ are strictly monotonic functions such that $h_{it}(j) \equiv h(\mu_{i}-\rho[z_{it}(j),e_{it}(j)])$ always varies when $e_{it}(j)$ does.}
 
The pair of histories $A$ and $B$ in equation \eqref{eq_example_1_histories_nozeros} is only one of the many history pairs with identification power for the structural parameters. For instance, we can extend this example by including periods of no purchase. Let $\textbf{0}_{n}$ represent a vector of $n$ zeros. For $k,j \geq 1$ with $k \neq j$, and any two natural numbers $n_{1}$ and $n_{2}$, consider the following choice histories:
\begin{equation}
    A 
    \text{ } = \text{ }
    \left(
        k, \text{ } 
        \textbf{0}_{n_{1}},  \text{ }
        j, \text{ }
        \textbf{0}_{n_{2}},  \text{ }
        k, \text{ }
        \textbf{0}_{n_{2}},  \text{ }
        j
    \right)
    \quad ; \quad
        B
    \text{ } = \text{ }
    \left(
        k, \text{ } 
        \textbf{0}_{n_{1}},  \text{ }
        k, \text{ }
        \textbf{0}_{n_{2}},  \text{ }
        j, \text{ }
        \textbf{0}_{n_{2}},  \text{ }
        j
    \right) 
\label{eq_example_1_histories_withzeros}
\end{equation}
It is straightforward to show that the difference between the log-probabilities of these two histories has the following expression:
\begin{equation}
    \begin{array}
    [c]{ccl}
    \log \mathbb{P}(A) - \log \mathbb{P}(B)
    & = &
    \displaystyle{\sum_{t=n_{1}+2}^{n_{1}+n_{2}+2}}
    v_{\boldsymbol{\alpha}_{i}}(j,\mathbf{z}_{t}) -
    v_{\boldsymbol{\alpha}_{i}}(k,\mathbf{z}_{t}) +
    \displaystyle{\sum_{t=n_{1}+n_{2}+3}^{n_{1}+2n_{2}+3}}
    v_{\boldsymbol{\alpha}_{i}}(k,\mathbf{z}_{t}) -
    v_{\boldsymbol{\alpha}_{i}}(j,\mathbf{z}_{t}) \\
    &  & \\
    & - &
    \displaystyle{\sum_{t=n_{1}+3}^{n_{1}+n_{2}+3}}
    \sigma_{\boldsymbol{\alpha}_{i}}(j,\mathbf{z}_{t},\mathbf{e}_{t}) -
    \sigma_{\boldsymbol{\alpha}_{i}}(k,\mathbf{z}_{t},\mathbf{e}_{t}) + 
    \displaystyle{\sum_{t=n_{1}+n_{2}+4}^{n_{1}+2n_{2}+4}}
    \sigma_{\boldsymbol{\alpha}_{i}}(k,\mathbf{z}_{t},\mathbf{e}_{t}) -
    \sigma_{\boldsymbol{\alpha}_{i}}(j,\mathbf{z}_{t},\mathbf{e}_{t}) \\    &  & \\
    & - &
    \widetilde{\beta}^{sc}(k,j) + 
    \gamma \text{ } 
    \left( 
        h_{n_{1}+2}(j) - h_{n_{1}+n_{2}+3}(j) - 
        h_{n_{1}+2}(k) + h_{n_{1}+n_{2}+3}(k)
    \right).
    \end{array}
\label{eq_logP_difference_noduration_withzeros}
\end{equation}
To eliminate the unobserved heterogeneity $\boldsymbol{\alpha}_{i}$ from this difference we need the following conditions on prices: (i) the permanent component $\mathbf{z}_{t}$ is constant from period $n_{1}+2$ to $n_{1}+2n_{2}+4$; and (ii) the transitory component $\mathbf{e}_{t}$ is constant between periods $n_{1}+3$ and $n_{1}+2n_{2}+4$. Under these conditions, we have that:
\begin{equation}
    \log \mathbb{P}(A) - \log \mathbb{P}(B) = 
    - \widetilde{\beta}^{sc}(k,j) 
    + \gamma \text{ } 
    \left( 
        h_{n_{1}+2}(j) - h_{n_{1}+3}(j) - h_{n_{1}+2}(k) + h_{n_{1}+3}(k)
    \right).
\label{eq_logP_difference_noduration_withzeros_short}
\end{equation}
This equation shows that a change between periods $n_{1}+2$ and $n_{1}+3$ in the transitory component of the price of product $j$ or $k$ identifies parameter $\gamma$. The switching cost parameter $\widetilde{\beta}^{sc}(k,j)$ is identified from histories where this transitory component is constant.

\subsection{Dynamic demand with duration dependence}

Consider now the demand model with duration dependence. The expression for the log-CCP function is:
\begin{equation}
    \log P_{it} = 
    \left\{
    \begin{array}
    [c]{lcl}
    \alpha_{i}(\ell_{it}) - \beta^{dep}(\ell_{it}) \text{ } d_{it} + 
    v_{\boldsymbol{\alpha}_{i}}(\ell_{it},d_{it}+1,\mathbf{z}_{t}) -
    \sigma_{\boldsymbol{\alpha}_{i}}\left(\ell_{it},d_{it},\mathbf{z}_{t},\mathbf{e}_{t}\right)
    & if & y_{it}=0
    \\
    \alpha_{i}(j) + \gamma \text{ } h_{it}(j)
    - \beta^{sc}(\ell_{it},j) +
    v_{\boldsymbol{\alpha}_{i}}(j,1,\mathbf{z}_{t}) -
    \sigma_{\boldsymbol{\alpha}_{i}}\left(\ell_{it},d_{it},\mathbf{z}_{t},\mathbf{e}_{t}\right) 
    & if & y_{it}=j>0  
    \end{array}
    \right. 
\label{eq_logCCP_withduration}
\end{equation}

First, it is straightforward to verify that the pair of histories in equation
\eqref{eq_example_1_histories_nozeros} still identifies the switching cost parameters $\widetilde{\beta}_{kj}^{sc}$ and the price sensitivity parameter $\gamma$. Therefore, I focus here on the identification of the depreciation parameters $\beta^{dep}(j)$. Furthermore, for notational simplicity, I consider here that the two price components are constant over the considered choice histories and omit $\mathbf{z}_{t}$ and $\mathbf{e}_{t}$ as arguments.

Let $n$ be a natural number such that $2 \leq n \leq (T-2)/2$. Consider the following choice histories, both with initial duration $d_{1}=1$:
\begin{equation}
    A 
    \text{ } = \text{ }
    \left(
        j, \text{ } 
        \textbf{0}_{n-1},  \text{ }
        j, \text{ }
        \textbf{0}_{n+1}
    \right)
    \quad ; \quad
        B
    \text{ } = \text{ }
    \left(
        j, \text{ } 
        \textbf{0}_{n},  \text{ }
        j, \text{ }
        \textbf{0}_{n}
    \right) 
\label{eq_example_withduration}
\end{equation}
Note that these histories do not contain any product switching event such that switching cost parameters do not appear in the probabilities of these choice histories. Taking into account the structure of the log-CCP in equation \eqref{eq_logCCP_withduration}, we have the following expressions for log-probabilities of these choice histories:
\begin{equation}
    \begin{array}
    [c]{ccl}
    \log \mathbb{P}(A)
    & = &
    \log p^{\ast}_{1}(j,d_{1} | \boldsymbol{\alpha}_{i}) +
    2(n+1)\alpha_{i}(j) -
    \displaystyle{\sum_{d=1}^{n-1}} \beta^{dep}(j) d -
    \displaystyle{\sum_{d=1}^{n+1}} \beta^{dep}(j) d \\
    & + &
    \displaystyle{\sum_{d=2}^{n}} v_{\boldsymbol{\alpha}_{i}}(j,d) +
    \displaystyle{\sum_{d=1}^{n+2}} v_{\boldsymbol{\alpha}_{i}}(j,d) -
    \displaystyle{\sum_{d=1}^{n}} \sigma_{\boldsymbol{\alpha}_{i}}(j,d) -
    \displaystyle{\sum_{d=1}^{n+1}} \sigma_{\boldsymbol{\alpha}_{i}}(j,d).
    \end{array}
\label{eq_logPA_withduration}
\end{equation}

\begin{equation}
    \begin{array}
    [c]{ccl}
    \log \mathbb{P}(B)
    & = &
    \log p^{\ast}_{1}(j,d_{1} | \boldsymbol{\alpha}_{i}) +
    2(n+1)\alpha_{i}(j) -
    \displaystyle{\sum_{d=1}^{n}} \beta^{dep}(j) d -
    \displaystyle{\sum_{d=1}^{n-1}} \beta^{dep}(j) d \\
    & + &
    \displaystyle{\sum_{d=2}^{n+1}} v_{\boldsymbol{\alpha}_{i}}(j,d) +
    \displaystyle{\sum_{d=1}^{n+1}} v_{\boldsymbol{\alpha}_{i}}(j,d) -
    \displaystyle{\sum_{d=1}^{n+1}} \sigma_{\boldsymbol{\alpha}_{i}}(j,d) -
    \displaystyle{\sum_{d=1}^{n}} \sigma_{\boldsymbol{\alpha}_{i}}(j,d).
    \end{array}
\label{eq_logPB_withduration}
\end{equation}
They imply the following expression for the difference between the log-probabilities:
\begin{equation}
    \log \mathbb{P}(A) - \log \mathbb{P}(B)
    \text{ } = \text{ } 
    - \beta^{dep}(j)  + 
    v_{\boldsymbol{\alpha}_{i}}(j,n+2) -
    v_{\boldsymbol{\alpha}_{i}}(j,n+1)
\label{eq_difference_logP_withduration}
\end{equation}
The expression in the right-hand-side still depends on the incidental parameters $\boldsymbol{\alpha}_{i}$ such that, without further restrictions, this pair of histories does not identify parameter $\beta^{dep}(j)$.

In general, without further restrictions, the depreciation rates $\beta^{dep}(j)$ are not identified in the forward-looking model.\footnote{Note that in a myopic model (i.e., $\delta_{i}=0$) all the continuation values $v_{\boldsymbol{\alpha}_{i}}(j,d)$ are zero, such that equation \eqref{eq_difference_logP_withduration} implies the identification of $\beta^{dep}(j)$ from $\log \mathbb{P}(A) - \log \mathbb{P}(B)$. Here, I do not impose this restriction and consider that consumers can be forward-looking.} To obtain identification of these parameters, I follow the same approach as \citeauthor{aguirregabiria_gu_2021} (\citeyear{aguirregabiria_gu_2021}) and impose the following restriction.

\bigskip

\begin{assumption}
    \textit{For any product $j$, there is a value of duration $d_{j}^{*}$ -- which can vary across products -- such that the structural function that captures the depreciation effect is:}
    \begin{equation}
        \left\{
        \begin{array}[c]{lcl}
            \beta^{dep}(j) \text{ } n 
            & if &
            n \leq d_{j}^{*} 
            \\ 
            \beta^{dep}(j) \text{ } d_{j}^{*} 
            & if &
            n \geq d_{j}^{*}. \qquad \blacksquare
        \end{array}
        \right.
\label{eq:restriction_depreciation}
\end{equation}
\label{assumption_3}
\end{assumption}

\bigskip

Importantly, as established in Proposition 6 in \citeauthor{aguirregabiria_gu_2021} (\citeyear{aguirregabiria_gu_2021}), the value of $d_{j}^{*}$ is identified from the data as long as it is not larger than $(T-1)/2$. I reproduce here this result from \citeauthor{aguirregabiria_gu_2021} (\citeyear{aguirregabiria_gu_2021}).

\bigskip

\noindent \textbf{PROPOSITION.} \textit{Let $d_{j}^{*}$ be the value of duration defined in \textcolor{blue}{Assumption} \ref{assumption_3} above. For any product $j$ and any duration $n$ with $2n + 1 \leq T$, define the pair of histories 
$A_{j,n} = (j, \boldsymbol{0}_{n-1}, j, \boldsymbol{0}_{n+1})$ and $B_{j,n} = (j, \boldsymbol{0}_{n}, j, \boldsymbol{0}_{n})$. If $d_{j}^{*} \leq (T-1)/2$, then the value $d_{j}^{*}$ is point identified from the following expression:}
\begin{equation}
    d_{j}^{*} \text{ } = \text{ }
    max \{ 
         n \text{ } : \text{ }
         \log \mathbb{P}(A_{j,n}) - \log \mathbb{P}(B_{j,n})
         \neq 0
    \}  \qquad \blacksquare
\end{equation}

\bigskip

An important implication of \textcolor{blue}{Assumption} \ref{assumption_3} is that the continuation value function is such that $v_{\boldsymbol{\alpha}_{i}}(j,n) = v_{\boldsymbol{\alpha}_{i}}(j,d_{j}^{*})$ for any duration $n \geq d_{j}^{*}$. Combining this property with equation \eqref{eq_difference_logP_withduration}, we have that for $n = d_{j}^{*}-1$:
\begin{equation}
    \log \mathbb{P}(A_{j,n}) - \log \mathbb{P}(B_{j,n})
    \text{ } = \text{ } 
    - \beta^{dep}(j) 
\label{eq_identification_betadep}
\end{equation}
such that the depreciation rate $\beta^{dep}(j)$ is identified.

\section{Estimation \label{sec_estimation}}

\subsection{Estimation of structural parameters \label{sec_estimation_parameters}}

Let $\boldsymbol{\theta}$ be the vector of structural parameters as defined in equation \eqref{eq:definition_theta}. The dataset is $\{y_{it}, \mathbf{z}_{it}, \mathbf{e}_{it}: i=1,2,...,N; t=1,2,...,T\}$. Let $\widetilde{\mathbf{z}}_{i}$ and $\widetilde{\mathbf{e}}_{i}$ the vectors with the time series of prices $\{\mathbf{z}_{it}: t=1,2,...,T\}$ and $\{\mathbf{e}_{it}: t=1,2,...,T\}$, respectively. 

In the identification results in Section \ref{sec_identification}, we represent a sufficient statistic for $\boldsymbol{\alpha}_{i}$ as a binary indicator that combines two conditions: the condition that choice history $\mathbf{y}_{i}$ is either equal to $A$ or to $B$; and a set of restrictions on the evolution of prices. For instance, for the identification of switching costs and price parameters, the persistent component $\mathbf{z}_{it}$ should be constant between periods $n_{1} + 2$ and $n_{1} + 2n_{2} + 4$, and the transitory component $\mathbf{e}_{it}$ should be constant between periods $n_{1}+3$ and $n_{1}+2n_{2}+4$. We represent these restrictions on prices using the system of equations $r(\widetilde{\mathbf{z_{i}}},\widetilde{\mathbf{e_{i}}})=\mathbf{0}$. Therefore, a sufficient statistic $s_{i}$ is the following binary indicator:
\begin{equation}
    s_{i} \equiv 
    s\left(
        \mathbf{y}_{i}, \widetilde{\mathbf{z}}_{i}, \widetilde{\mathbf{e}}_{i}
    \right) =
    1\{ 
        \mathbf{y}_{i} \in A \cup B \text{ and } 
        r(\widetilde{\mathbf{z}}_{i},\widetilde{\mathbf{e}}_{i})=\mathbf{0}
    \}
\end{equation}
We have shown that:
\begin{equation}
    \mathbb{P}(\mathbf{y}_{i}|\widetilde{\mathbf{z}}_{i},\widetilde{\mathbf{e}}_{i},s_{i}=1) =
    \frac{ 
        \exp\{ 
        c(\mathbf{y}_{i},\widetilde{\mathbf{z}}_{i},\widetilde{\mathbf{e}}_{i})
        ^{\prime} \boldsymbol{\theta}    
        \}
    }
    {
        \exp\{ 
        c(A,\widetilde{\mathbf{z}}_{i},\widetilde{\mathbf{e}}_{i})^{\prime} \boldsymbol{\theta}
        \} +
        \exp\{ 
        c(B,\widetilde{\mathbf{z}}_{i},\widetilde{\mathbf{e}}_{i})^{\prime} \boldsymbol{\theta}
        \}
    }
\end{equation}
where $c(\mathbf{y}_{i},\widetilde{\mathbf{z}}_{i},\widetilde{\mathbf{e}}_{i})$ is a vector of known statistics. 

There are many pairs of histories $A$ and $B$ that provide sufficient statistics for $\boldsymbol{\alpha}_{i}$ and have identification power for $\boldsymbol{\theta}$. Let these sufficient statistics be indexed by $m \in \{1,2, ..., M\}$. We also index by $m$ the different elements that define sufficient statistic $s_{i}^{m}$, that is: the corresponding pair of choice histories, $(A^{m},B^{m})$; the restrictions on prices, $r^{m}(\widetilde{\mathbf{z}}_{i},\widetilde{\mathbf{e}}_{i})=\mathbf{0}$; and the identifying statistics, $c^{m}(\mathbf{y}_{i},\widetilde{\mathbf{z}}_{i},\widetilde{\mathbf{e}}_{i})$. Note that we can use $m$ also to index different sub-periods in the panel dataset.\footnote{For instance, we can divide a panel dataset with $T=12$ periods into three different sub-panels with four time periods each, and construct sufficient statistics separately for each sub-panel.} Given these $M$ sufficient statistics, we can define the following \textit{conditional log-likelihood function}:
\begin{equation}
\begin{array}[c]{l}
    \mathcal{L}(\boldsymbol{\theta}) = \\
    \displaystyle{\sum_{m=1}^{M}} \displaystyle{\sum_{i=1}^{N}}
    1\{ \mathbf{y}_{i} \in A^{m} \cup B^{m} \} \text{ }
    1\{r^{m}(\widetilde{\mathbf{z}}_{i},\widetilde{\mathbf{e}}_{i})=\mathbf{0}\} 
    \text{ }
    \log \left(
       \frac{ 
        \exp\{ 
        c^{m}(\mathbf{y}_{i},\widetilde{\mathbf{z}}_{i},\widetilde{\mathbf{e}}_{i})
        ^{\prime} \boldsymbol{\theta}    
        \}
    }
    {\exp\{ 
          c^{m}(A^{m},\widetilde{\mathbf{z}}_{i},\widetilde{\mathbf{e}}_{i})^{\prime} \boldsymbol{\theta}
        \} +
        \exp\{ 
            c^{m}(B^{m},\widetilde{\mathbf{z}}_{i},\widetilde{\mathbf{e}}_{i})^{\prime} \boldsymbol{\theta}
        \}
    }
    \right)        
\end{array}
\end{equation}
The CML estimator is the value of $\boldsymbol{\theta}$ that maximizes $\mathcal{L}(\boldsymbol{\theta})$, which is a globally concave function.

In many possible applications, such as those using weekly or daily supermarket scanner data, price stickiness implies such that the restrictions  $r^{m}(\widetilde{\mathbf{z}}_{i},\widetilde{\mathbf{e}}_{i})=\mathbf{0}$ hold for a non-negligible fraction of observations. Nevertheless, imposing exactly these restrictions typically implies losing a substantial amount of observations. This is exactly the issue that motivates the \textit{Kernel Weighted CML} in \citeauthor{honore_kyriazidou_2000} (\citeyear{honore_kyriazidou_2000}). In the log-likelihood function, we replace the indicator $1\{r^{m}(\widetilde{\mathbf{z}}_{i},\widetilde{\mathbf{e}}_{i})=\mathbf{0}\}$ with a weight that depends inversely on the magnitude of vector $| r^{m}(\widetilde{\mathbf{z}}_{i},\widetilde{\mathbf{e}}_{i})|$ such that we put more weight in the log-likelihood of observations for which $| r^{m}(\widetilde{\mathbf{z}}_{i},\widetilde{\mathbf{e}}_{i})|$ is close to zero. More specifically, we consider the following Kernel Weighted conditional log-likelihood function:
\begin{equation}
\begin{array}[c]{l}
    \mathcal{L}^{KW}(\boldsymbol{\theta}) = \\
    \displaystyle{\sum_{m=1}^{M}} \displaystyle{\sum_{i=1}^{N}}
    1\{ \mathbf{y}_{i} \in A^{m} \cup B^{m} \} \text{ }   
    K\left(
        \frac{r^{m}(\widetilde{\mathbf{z}}_{i},\widetilde{\mathbf{e}}_{i})}
        {b_{N}}
    \right)
    \text{ }
    \log \left(
       \frac{ 
        \exp\{ 
        c^{m}(\mathbf{y}_{i},\widetilde{\mathbf{z}}_{i},\widetilde{\mathbf{e}}_{i})
        ^{\prime} \boldsymbol{\theta}    
        \}
    }
    {\exp\{ 
          c^{m}(A^{m},\widetilde{\mathbf{z}}_{i},\widetilde{\mathbf{e}}_{i})^{\prime} \boldsymbol{\theta}
        \} +
        \exp\{ 
            c^{m}(B^{m},\widetilde{\mathbf{z}}_{i},\widetilde{\mathbf{e}}_{i})^{\prime} \boldsymbol{\theta}
        \}
    }
    \right)        
\end{array}
\end{equation}
where $K(.)$ is a kernel density function that satisfies the regularity condition $K(v) \rightarrow 0$ as $||v|| \rightarrow \infty$, and $b_{N}$ is a bandwidth parameter such that $b_{N} \rightarrow 0$ and $N b_{N} \rightarrow \infty$ and $N \rightarrow \infty$. As shown by \citeauthor{honore_kyriazidou_2000} (\citeyear{honore_kyriazidou_2000}), this estimator is consistent and asymptotically normal. If prices $\mathbf{z}_{it}$ and $\mathbf{e}_{it}$ have discrete support, the rate of convergence of the estimator is $N^{-1/2}$. Otherwise, with continuous prices, the rate of convergence is slower than $N^{-1/2}$.

\subsection{Average marginal effects and counterfactual experiments for aggregate consumer demand \label{sec:estimation_counterfactuals}}

An important motivation for the estimation of structural demand models is using them to obtain \textit{Average Marginal Effects (AMEs)} and \textit{counterfactual experiments} that provide the causal effect on aggregate consumer demand of hypothetical changes in structural parameters or in state variables. In the context of dynamic demand models, short-run and long-run aggregate demand elasticities are important examples of \textit{AMEs}. Possible counterfactual experiments include evaluating the effect on aggregate consumer demand of removing sources of dynamics (e.g., making $\widetilde{\beta}^{sc}$ or $\beta^{dep}$ equal to zero), or introducing hypothetical new products.\footnote{Some of these counterfactuals may require endogenizing firms' pricing decisions, but, as a first step, they can be implemented keeping prices fixed.}

In Fixed Effect discrete choice models with short panels, the distribution of the time-invariant unobserved heterogeneity is not identified. This is because the data consist of a finite number of probabilities -- as many as the number of possible choice histories -- but the distribution of the unobserved heterogeneity has infinite dimension. Since aggregate consumer demand is an expectation over the
distribution of the unobserved heterogeneity, and this distribution is not identified, the common wisdom is that Fixed Effect approaches cannot (point) identify AMEs or counterfactuals on aggregate demand. Here I present a brief discussion of two possible approaches.

\bigskip

\noindent \textbf{(i) Random Effects approach.} After the Fixed Effects estimation of the structural parameters $\boldsymbol{\theta}$, the researcher can consider a parametric Random Effects model for the distribution of the unobserved heterogeneity, $f_{\boldsymbol{\alpha}}(\boldsymbol{\alpha}_{i})$, and the probability function for the initial conditions, $p^{*}(\ell_{i1},d_{i1}|\boldsymbol{\alpha}_{i})$. The parameters in these two functions can be estimated using full maximum likelihood (or other full solution method), taking as given the 
estimate of $\boldsymbol{\theta}$. Given the Fixed Effects estimate of $\boldsymbol{\theta}$ and the Random Effects estimate of the parameters in $f_{\boldsymbol{\alpha}}(\boldsymbol{\alpha}_{i})$ and $p^{*}(\ell_{i1},d_{i1}|\boldsymbol{\alpha}_{i})$, the researcher can calculate any AME or counterfactual experiment.

Though conceptually simple, this approach suffers from two important limitations. A first issue is the potential misspecification of the parametric restrictions imposed on functions $f_{\boldsymbol{\alpha}}(\boldsymbol{\alpha}_{i})$ and $p^{*}(\ell_{i1},d_{i1}|\boldsymbol{\alpha}_{i})$. Though the Fixed Effect estimation of $\boldsymbol{\theta}$ is robust to this misspecification, the estimates of AMEs and counterfactuals are not, as they depend on the distribution of the unobservables. A second issue is that the full solution estimation of the model requires computing continuation value functions, that may be impractical in applications when the number of products is not small.

\bigskip

\noindent \textbf{(ii) Recent results on identification of AMEs in Fixed Effects discrete choice models.} Here I briefly describe recent developments on identification and estimation of AMEs and counterfactuals in Fixed Effects discrete choice models with short panels. In contrast to the Random Effects approach in item (i) above, these methods share the same attractive features of robustness and computational simplicity in the Fixed Effects estimation of $\boldsymbol{\theta}$. However, this literature is very recent and the results are still limited. 

\citeauthor{chernozhukov_fernandez_2013} (\citeyear{chernozhukov_fernandez_2013}) study the identification and estimation of AMEs and counterfactuals in Fixed Effects binary choice models. The AMEs of interest consist of the causal effect on aggregate demand (i.e., CCP function integrated  over the distribution of unobserved heterogeneity) of a change in an explanatory variable for all the consumers (e.g., a change in market prices) or a change in structural parameters (e.g., making $\beta^{dep}$ equal to zero). The authors characterize the identified set of the AME, and propose an algorithm to estimate this set.

For Fixed Effects dynamic multinomial logit models as the ones in this paper, \citeauthor{aguirregabiria_carro_2021} (\citeyear{aguirregabiria_carro_2021}) show the point identification of AMEs that consist of a change in the dynamic explanatory variables, i.e., a change in product choice in last purchased ($\ell_{it}$), or in duration since last purchased ($d_{it}$). Their identification results apply to the model in this paper. Despite the distribution of the unobserved heterogeneity is not identified, they show that these AMEs are relatively simple expressions of the probability of some choice histories and of structural parameters. In a similar spirit, 
\citeauthor{davezies_2021} (\citeyear{davezies_2021}), and \citeauthor{pakel_weidner2021} (\citeyear{pakel_weidner2021}) present new results on set identification of a broader class of AMEs in Fixed Effects discrete choice models. They also propose simple algorithms to compute the identified sets.


\section{Conclusions \label{sec_conclusions}}

This paper presents a Fixed Effects dynamic panel data model of demand for different products where consumers are forward looking. I apply and extend recent results from \citeauthor{aguirregabiria_gu_2021} (\citeyear{aguirregabiria_gu_2021}) to establish the identification of all structural parameters in this model. Several extensions of the results in this paper are interesting topics for further research.

First, the specification of the sources of demand dynamics in the model of this paper is restrictive. 
For instance, as mentioned in Section \ref{sec:model_utility}, more realistic models of habit formation include habit stock variables for each product, where the stock increases when the product is purchased and declines otherwise. Also, the model does not accommodate some forms of consumer learning for experienced goods used in the literature with Random Effects models (e.g., \citeauthor{ching_2010}, \citeyear{ching_2010}).

Second, the model assumes that consumers buy at most one unit of the product per period. However, it is well-known that forward-looking consumers can buy for inventory (\citeauthor{hendel_nevo_ecma_2006}, \citeyear{hendel_nevo_ecma_2006}, \citeyear{hendel_nevo_aer_2013}). It would be interesting to extend the model to incorporate the possibility of consumers purchasing multiple units.

Third, as mentioned in Section \ref{sec:estimation_counterfactuals}, an important motivation for the estimation of structural demand models is the evaluation of counterfactual experiments on aggregate consumer demand. Developments on (point or set) identification and estimation of this type of counterfactuals would be very useful in this literature.

Fourth, in the same spirit as the the recent work by \citeauthor{mysliwski_sanches_2020}, (\citeyear{mysliwski_sanches_2020}), this dynamic demand can be combined with a dynamic game of price competition. \textcolor{blue}{Assumption} \ref{assumption_1} on the two price components -- persistent and transitory -- is consistent with an equilibrium of a dynamic pricing game with i.i.d. firms' private information.

\clearpage

\baselineskip14pt

\bibliographystyle{econometrica}

\bibliography{references}

\end{document}